\documentclass[12pt, reqno, a4paper,oneside]{amsart}

\rmfamily{\fontsize{12pt}{\baselineskip}\selectfont}
 
\usepackage{graphicx}
\usepackage{amsfonts, amsmath, amssymb, amsbsy, amsthm}
\usepackage{bm}
\usepackage{color}
\usepackage{xcolor}
\usepackage{booktabs} 
\usepackage{hyperref}
\usepackage{mathrsfs}
\usepackage{longtable}
\usepackage[top=2.5cm, bottom=2.0cm, left=2.0cm, right=2.0cm]{geometry}
\usepackage{epstopdf}
\usepackage{stmaryrd}
\usepackage[normalem]{ulem}
\usepackage{cancel}
\usepackage{scalerel}
\usepackage{enumitem}
\usepackage{algorithm}
\usepackage{cleveref}
\usepackage{subcaption}
\usepackage{algorithm, algorithmicx}
\usepackage[noend]{algpseudocode}
\usepackage{lineno}
\usepackage{enumitem}
\usepackage{diagbox}
\usepackage{multirow}
\renewcommand{\algorithmiccomment}[1]{\bgroup\hfill//~#1\egroup}

\usepackage{environments}
\numberwithin{theorem}{section}
\numberwithin{equation}{section}

\definecolor{yscol}{HTML}{6622AA}

\title[Many-Body Coarse-Grained Molecular Dynamics with ACE]{Many-Body Coarse-Grained Molecular Dynamics with the Atomic Cluster Expansion}

\author{Yangshuai Wang}
\address{Yangshuai Wang\\
Department of Mathematics\\
National University of Singapore\\
10 Lower Kent Ridge Road\\
Singapore
}
\email{yswang@nus.edu.sg}

\author{Gabor Csanyi}
\address{Gabor Csanyi \\
Department of Engineering \\
University of Cambridge \\
Trumpington Street \\ 
Cambridge CB2 1PZ\\
United Kingdom
}
\email{gc121@cam.ac.uk}

\author{Christoph Ortner}
\address{Christoph Ortner\\
Department of Mathematics\\
University of British Columbia\\
1984 Mathematics Road\\
Vancouver, British Columbia\\
Canada
}
\email{ortner@math.ubc.ca}

\date{\today}

\begin{document}
\maketitle

\begin{abstract}
Molecular dynamics (MD) simulations provide detailed insight into atomic-scale mechanisms but are inherently restricted to small spatio-temporal scales. Coarse-grained molecular dynamics (CGMD) techniques allow simulations of much larger systems over extended timescales. 
In theory, these techniques can be quantitatively accurate, but common practice is to only target qualitatively correct behaviour of coarse-grained models. Recent advances in applying machine learning methodology in this setting are now being applied to create also quantitatively accurate CGMD models.
We demonstrate how the Atomic Cluster Expansion parameterization (Drautz, 2019)  can be used in this task to construct highly efficient, interpretable and accurate CGMD models. We focus in particular on exploring the role of many-body effects.
%
\end{abstract}

\def\<{\langle}
\def\>{\rangle}

\def\a{{\bf a}}
\def\z{{\bf z}}
\def\q{{\bf q}}
\def\p{{\bf p}}
\def\b{{\bf b}}
\def\r{{\bf r}}
\def\f{{\bf f}}
\def\v{{\bf v}}

\def\bxi{{\bf \xi}}

\def\Z{{\bf Z}}
\def\Q{{\bf Q}}
\def\P{{\bf P}}
\def\M{{\bf M}}
\def\K{{\bf K}}
\def\V{{\bf V}}
\def\R{{\bf R}}
\def\F{{\bf F}}
\def\G{{\bf G}}
\def\bzeta{{\bf \zeta}}

\def\Up{\Upsilon}

\section{Introduction}
\label{sec:intro}
Molecular dynamics (MD) simulations offer insights into atomic-level processes but are inherently limited in their ability to efficiently capture phenomena across larger spatio-temporal scales. Bridging the gap between micro-scale atomic resolution and meso-scale phenomena remains a significant challenge in understanding complex systems~\cite{van2020roadmap, jin2022bottom}. Coarse-graining (CG) techniques are a promising tool to extend simulation capabilities to larger systems and longer timescales. These techniques work by grouping atoms into larger particles, allowing the focus to shift from detailed atomic interactions to broader, system-level behaviors. The resulting coarse-grained molecular dynamics (CGMD) methods have been widely applied and extensively studied in a wide variety of application scenarios~\cite{bond2007coarse, hsu2019coarse, jin2020temperature, de2013improved, kmiecik2016coarse, marrink2013perspective, rudzinski2021dynamical, wang2009effective}.

In CGMD, accurately representing the potential of mean force (PMF) is crucial for capturing system behavior~\cite{kmiecik2016coarse, pagonabarraga2001dissipative, reith2003deriving}. 
Therefore, the precise and efficient description of the PMF is a key challenge in this field. 
Due to many parallels between modeling interatomic potentials (or, force fields) and the PMF, advances in machine learning interatomic potentials (MLIPs)~\cite{2019-ship1, Bart10, behler07, Drautz19, Shapeev16} can be leveraged for this task. Works applying MLIP techniques to model the PMF include~\cite{durumeric2024learning, duschatko2024thermodynamically, husic2020coarse, john2017many, loose2023coarse, majewski2023machine, patra2019coarse, wang2019machine, zhang2018deepcg}. We refer to the recent review~\cite{sahrmann2025emergence} for a more comprehensive discussion.

In the same way as interatomic interactions, the interaction between coarse-grained particles also involve many-body effects that cannot always be ignored~\cite{han2021constructing, wang2021multi, zaporozhets2023multibody}. While the deep learning techniques cited above do incorporate some many-body effects, a systematic, efficient and interpretable approach for non-bonded many-body terms remains an ongoing challenge in developing accurate CG models. In the field of MLIPs, extensive research has focused on this aspect. Of particular note are the Moment Tensor Potential~\cite{Shapeev16} and the Atomic Cluster Expansion (ACE)~\cite{Drautz19} methods which provide systematic and flexible but computationally efficient solutions. Although the ACE method and its extensions~\cite{2019-ship1, batatia2022mace, Drautz19, lysogorskiy2021performant, wang2024theoretical} has been widely used to model interatomic interactions, to the best of our knowledge, it has not been applied to approximate the PMF in the CGMD context.

In this work, we investigate the use of the ACE method to construct a computationally efficient many-body coarse-grained ``ACE-CG'' model.
Traditionally, constructing data-driven CG models involves matching mean forces, a process typically requiring computationally intensive constrained or restrained simulations~\cite{john2017many, wang2019machine, zhang2018deepcg}. To overcome this difficulty we utilize the force-matching scheme to optimize the ACE-CG model parameters~\cite{davtyan2015dynamic, izvekov2005multiscale}. In this scheme one directly derives free energy gradients from standard atomic forces (often called ``instantaneous forces'' in the CGMD literature) sampled from reference molecular dynamics trajectories. We test the ACE-CG method on two prototype systems: star polymers and methanol fluids. We demonstrate the ACE-CG models' capability to accurately represent equilibrium properties such as radial distribution functions (RDFs) and angular distribution functions (ADFs). 
Furthermore we demonstrate the importance of including many-body effects to improve qualitative and/or quantiative accuracy of the predicted equilibrium properties, compared to pairwise potentials. Our results demonstrate the capabilities of the ACE formalism in the CGMD context, achieving systematically improvable accuracy at high computational efficiency. 


The scope of this work is limited to the prediction of equilibrium properties determined by the conservative potential function of a set of extensive CG variables. For the conformational free energy of non-extensive CG variables, several machine learning-based approaches have been developed~\cite{lemke2017neural, mones2016exploration, stecher2014free, wu2023k, zavadlav2018multiscale}; see also a recent review~\cite{noe2020machine} and the references therein. Moreover, to accurately predict dynamic properties, it is necessary to properly introduce memory and coherent noise terms arising from unresolved variables into the CG model~\cite{hijon2010mori, klippenstein2021introducing, lei2016data, lyu2023construction}. These are limitations still to be addressed to establish ACE-CG as a general-purpose framework for CGMD simulation. 


%

\section{Models and Methods}
\label{sec:method}
Previous works~\cite{john2017many, wang2019machine, zhang2018deepcg} explored the use of machine learning methods to construct coarse-grained (CG) potentials. We provide a review of the general approach, the force matching principle, and then show how the framework can be adapted to develop a systematic and efficient many-body coarse-grained model using the Atomic Cluster Expansion (ACE).

\subsection{Consistent coarse-grained models}
\label{sec:sub:basicsCG}
A key objective in coarse-graining is to ensure that the equilibrium distribution of a system under a CG model matches that of the reference atomistic model. This principle aligns with the concept of {\it consistent} CG models~\cite{noid2013perspective, noid2008multiscale}. 
Preserving the equilibrium distribution enables consistent CG models to accurately describe equilibrium properties. In principle one can extend the approach to non-equilibrium properties such as time correlation functions~\cite{hijon2010mori, lu2014exact, teza2020exact}, but this is beyond the scope of the present work.

We consider an atomistic system with $n$ atoms with positions $\r_i$, momenta $\p_i$,
\begin{equation*}
    \r = \{\r_1, \ldots, \r_n\}, \quad \p = \{\p_1, \ldots, \p_n\},
\end{equation*}
as well as species $z_i$ (e.g. atomic number) and masses $m_i$. 
We assume that atoms are in the canonical (NVT) ensemble in equilibrium, under a potential energy $u(\{\r_i, z_i\}_{i=1}^n)$. Implicit in the definition is a computational cell, and we allow arbitrary length $n$ input configurations. 

Analogously, the state of the system in the CG model is defined by positions $\R_I$, momenta $\P_I$ of $N$ CG variables,
\begin{equation*}
    \R = \{\R_1, \ldots, \R_N\}, \qquad \P = \{\P_1, \ldots, \P_N\},
\end{equation*}
as well as {\em effective species} $\zeta_I$ and {\em effective masses} $M_I$. As is common, we use upper-case variables to refer to the CG model and lower-case variables to refer to the atomistic model (except for $\zeta_I$ to avoid confusion with common usage of $Z_I$ as atomic number). We assume that the two descriptions are explicitly connected via projections $\{\mathcal{M}_1, \ldots, \mathcal{M}_N\}$ such that $\R_I = \mathcal{M}_I(\r)$. 
The projections are many-to-one mappings: many atomistic configurations are mapped to the same CG configuration. 
For the sake of simplicity of presentation we assume that the projections are linear in the position, 
$\R_I = \mathcal{M}_I(\r) 
        = \sum_{i \in \mathcal{J}_I} w_{Ii}  \r_i,$
with $\sum_{i} w_{Ii} = 1$, where $\mathcal{J}_I$ denotes the set of atoms that form the CG particle $I$. The center of mass projection is particularly prevalent throughout the literature; it can be expressed as 
\begin{equation} \label{eq:ctr_mass}    
    w_{Ii} = \begin{cases}
        \frac{m_i}{\sum_{j \in \mathcal{J}_I} m_j}, & \text{if } i \in \mathcal{I}_I, \\ 
        0, & \text{otherwise}. 
    \end{cases} 
\end{equation}
Much more general choices are possible~\cite{voth2008coarse}. 

We require that our coarse-grained model can be used in a size-extensive way, hence it is crucial to define the effective coarse-grained species $\zeta_I$. For the sake of simplicity we define $\zeta_I$ to be the multi-set of species from which it is made up, i.e., 
\begin{equation} \label{eq:defn_mu}
    \zeta_I = \{z_i\}_{i \in \mathcal{J}_I}. 
\end{equation}
This definition is sufficient for our examples we present below, but it may be insufficient in general, e.g., local bonding topology should play a role in some cases. 

With the definition of $\zeta_I$ we then require that the weights $w_{Ii}$ can be written as 
\begin{equation}
    w_{Ii} = \mathcal{W}(\zeta_I; \{ z_i, m_i \}_{i \in \mathcal{J}_I} ), 
\end{equation}
which guarantees that the projection operator is again size-extensive.
This is clearly the case for the center of mass coarse-graining operator \eqref{eq:ctr_mass}.



\def\zz{{\bf z}}
\def\bzeta{{\bm \zeta}}

In the following it becomes notationally convenient to identify $(\r, \zz) = \{\r_i, z_i\}_{i=1}^n$ and $(\R, \bzeta) = \{\R_I, \zeta_I\}_{I = 1}^N$. A {\it consistent} CG potential $U(\R, \bzeta)$  
can be explicitly written as a {\it conditional free energy}, 
\begin{align}\label{eq:many-body-PMF-U}
   z(\r, \zz) =&~e^{-u(\r, \zz)/(k_B T)}, \nonumber \\
    Z(\R, \bzeta) =& \int z(\r, \zz) \prod^N_{I=1}\delta\big(\mathcal{M}_I(\r)-\R_I\big) {\rm d}\r,  \\ 
    U(\R, \bzeta) =& -k_B T \log Z(\R, \bzeta) + c, \nonumber
\end{align}
where $c$ is an arbitrary constant which we omit in the following, $k_B T$ is the thermal energy 
and $Z$ is the constrained partition function (as a function of the CG coordinates and CG species). See References~\cite{sprik1998free, stoltz2010free} and Appendix~\ref{sec:apd:preliminaries} for the details.




Numerous techniques have been developed to parameterize the CG potential $U$ in practice \cite{koehler2023flow, noid2008multiscale, shireen2022machine, thaler2022deep}.
The procedure can be divided into two steps: (1) the choice of an appropriate CG potential representation, and (2) the optimization of the parameters that define the potential representation. 
We will employ the ACE method~\cite{2019-ship1,Drautz19} for representing $U$ and the multiscale coarse-graining, or, force-matching method for parameter estimation. These methods are reviewed in the following sections.


\subsection{Atomic Cluster Expansion Parameterisation}
\label{sec:sub:sub:params}
Having written the CG potential $U$ in a size-extensive way, we will use a parameterization of $U$ that is able to accommodate this, as well as exploit the resulting symmetries: invariance under relabeling and under rotations and reflections. The Atomic Cluster Expansion (ACE) method has emerged as an efficient, systematic, interpretable and performant approach to accurately model high-dimensional many-particle interaction of this kind~\cite{2019-ship1, batatia2022mace, Drautz19, kovacs2021linear, lysogorskiy2021performant}. 
We introduce its use for representing the CG potential $U$. The construction is analogous the construction for machine learning interatomic potentials (MLIPs) to represent the potential energy $u$. We emphasize from the outset, though, that there are many alternative architectures that could be employed instead of ACE. Our reason to choose the ACE model is ease of use, and the straightforward manner in which we can explore the needed model complexity, in particular the need to incorporate many-body effects. 


\def\bth{{\bm \theta}}

We write the parameterized ACE-CG potential as $U^{\rm ACE}(\bth; \R, \bzeta)$, where $\bth$ is a parameter array. The first step is to decompose the total potential into site potentials,  
\begin{eqnarray}\label{eq:acecg_local_site}
    U^{\rm ACE}(\bth; \R, \bzeta) = \sum^{N}_{I=1} U^{\rm ACE}_{I}(\bth; \R, \bzeta).
\end{eqnarray}
The definition of $U^{\rm ACE}_{I}$ involves specifying a cutoff radius $R_{\rm cut}$ so that only CG particles $\R_J$ within distance $R_{IJ} < R_{\rm cut}$ contribute to the site energy $U^{\rm ACE}_{I}$. As far as we are aware there exists no theoretical justification for such a locality of CG interaction, but empirical studies support it~\cite{john2017many}. 

The second step is to express each site-potential in terms of a truncated {\it many-body expansion}~\cite{2019-ship1, Drautz19},  
\begin{eqnarray}\label{eq:acecg_local}
    U^{\rm ACE}_I(\bth; \R, \bzeta) = \sum_{\nu = 0}^{\nu_{\rm max}} 
        \sum_{1 \leq I_1 < \dots < I_\nu \leq N}
        W^{(\nu)}\big(\bth; \zeta_I, \R_{I_1}, \zeta_{I_1}, \dots, \R_{I_\nu}, \zeta_{I_\nu}). 
\end{eqnarray}
Each potential $W^{(\nu)}$ is then expanded in terms of a tensor product basis employing an irreducible representation of $O(3)$. One then performs a sequence of algebraic transformations to circumvent the combinatorial scaling of the many-body expansion and projects the resulting representation to a new basis that is invariant under rotations, reflections and relabeling of particles. The details of this procedure can be found in \cite{2019-ship1, Drautz19} and are summarized in Appendix~\ref{sec:apd:ACE}. 
It results in a linear model 
\begin{eqnarray}\label{eq:acecg}
    U^{\rm ACE}(\bth; \R, \bzeta) = \bth \cdot {\bm B}(\R, \bzeta), 
\end{eqnarray}
where ${\bm B}$ encodes the aforementioned symmetries and many-body expansion. Each parameter $\theta_j$  can be directly interpreted as a polynomial coefficient for one of the potentials $W^{(\nu)}$. Despite the underlying many-body expansion, the computational cost of $U^{\rm ACE}, \nabla U^{\rm ACE}$ scale linearly with the number of CG particles $N$ and linearly with the number of CG particles in each particle neighborhood.



\subsection{Force matching}
\label{sec:sub:sub:insF}
The CG force on a coarse site $I$ is given by the mean force 
\[
F_I(\R, \bzeta) = -\nabla_{\R_I} U(\R, \bzeta) 
\]
which can in principle be directly measured from constrained MD simulations~\cite{barth1995algorithms, sprik1998free}. Having direct access to mean force data suggests defining the loss function 
\begin{eqnarray}\label{eq:Lorigin}
 \mathcal{L}^{\rm MF}(\bth) = \sum_{(\R,\bzeta) \in \mathfrak{D}_{\rm CG}} \big|\nabla{U}^{\rm ACE}(\bth; \R, \bzeta) + F( \R, \bzeta)\big|^2,   
\end{eqnarray}
where the data set $\mathfrak{D}_{\rm CG}$ can be an arbitrary collection of coarse configurations $\R$, for example obtained from unconstrained MD or Monte Carlo simulations of the microscopic atomistic model and then applying the CG mapping $\mathcal{M}$. 

However, constrained MD simulations (cf.~Appendix~\ref{sec:apd:preliminaries}) are challenging, hence one often applies an alternative scheme where the instantaneous collective force (ICF) is employed instead~\cite{john2017many, wang2009effective}.
The ICF, denoted by $\mathcal{F}_I$ can be expressed in terms of the atomic forces $\f_i = -\nabla_{\r_i} u(\r, \zz)$ and the projectors $\mathcal{M}_I$.  
Since we assumed linear mappings, the ICF on a CG site $I$ becomes simply a weighted sum of the atomic forces,
\begin{eqnarray*}
    \mathcal{F}_I(\r, \zz) = \sum_{i} w_{Ii} \f_i(\r, \zz). 
\end{eqnarray*}
The ICF is identical to the local mean force employed in the adaptive biasing force method~\cite{stoltz2010free}.

Although the ICF is a deterministic function of the atomic coordinates $\r$, it can be viewed as a noisy sample of the mean force $F_{I}(\R, \bzeta)$. The connection is that the mean force is the conditional canonical average of the ICF (see Appendix~\ref{sec:apd:preliminaries}) 
\begin{eqnarray}\label{eq:cons_cg_forces}
    F_I(\R,\bzeta) =
    \frac{\int e^{-u(\r,\zz)/(k_B T)}\prod_{J}\delta\big(\mathcal{M}_J(\r)-\R_J\big) \mathcal{F}_I(\r, \zz) {\rm d}\r  }{\int e^{-u(\r, \zz)/(k_B T)}\prod_J\delta\big(\mathcal{M}_J(\r)-\R_J\big) {\rm d}\r }, 
\end{eqnarray}
where $\bzeta = \bzeta(\zz)$ is determined by the microscopic configuration for which we assume in this integral that $\zz$ remains constant. 
%

\label{sec:sub:sub:estimate}

The mean force loss function \eqref{eq:Lorigin} can now be replaced with 
\begin{eqnarray}\label{eq:Lins}
 \mathcal{L}(\bth) = \sum_{(\r, \zz) \in \mathfrak{D}_{\rm MD}}  \big|\nabla_{\r}{U}^{\rm ACE}(\bth; \mathcal{M}(\r), \bzeta(\zz)) + \mathcal{F}(\r, \zz)\big|^2.   
\end{eqnarray}
The training dataset $\mathfrak{D}_{\rm MD}$ now consistent of full-atom configurations. Unlike for \eqref{eq:Lorigin} it is now crucial that it respects the equilibrium distribution of the atomistic model. Employing the fact that the instantaneous force can be viewed as the mean force plus a random force $\mathcal{R}(\r)$ that depends on the microscopic configuration $(\r, \zz)$, it was shown in \cite{zhang2018deepcg} that the loss functions \eqref{eq:Lorigin} and \eqref{eq:Lins} have the same minimizer with respect to $\bth$. This equivalence justifies the usage of $\mathcal{L}(\bth)$ as the loss function.




\section{Results}
\label{sec:numer}
%
We evaluate the effectiveness of ACE-CG on two examples: methanol fluids and a star-polymer systems. In addition to demonstrating the feasibility and accuracy of the ACE-CG method, we will highlight many-body effects in the CG potential.

\subsection{General setup}
All simulations will be performed in the canonical (NVT) ensemble, utilizing  Langevin dynamics to maintain constant temperature, 
\begin{eqnarray}\label{eq:langevin}
    \frac{{\rm d}\hat{\mathbf{p}}}{{\rm d}t} = -\nabla \widehat{U}(\mathbf{r}) - \gamma \hat{\mathbf{p}} + \sqrt{2 \gamma k_B T} \, \boldsymbol{\eta}(t),
\end{eqnarray}
where $\gamma$ is the friction coefficient, and $\boldsymbol{\eta}(t)$ is a white noise vector. It is important to note that in our numerical experiments, we set $\nabla\widehat{U} = \nabla_{\mathbf{r}} u(\mathbf{r}, \mathbf{z})$ for the reference all-atom simulations and $\hat{\mathbf{p}}=\mathbf{p}$ is the corresponding momenta, while $\nabla\widehat{U} = \nabla_{\mathbf{R}} U^{\rm ACE}(\bth; \mathbf{R}, \boldsymbol{\zeta})$ is used for CGMD and also $\hat{\mathbf{p}}=\mathbf{P}$ for CG momenta. 

Typical equilibrium properties analyzed in CGMD include the radial and angular distribution functions (RDF and ADF), which serve as basic metrics for evaluating the quality of a coarse-grained model. For simplicity of presentation, we use lower-case variables to define these equilibrium properties, noting that the definitions for coarse-grained variables follow in a similar manner.

The RDF $g_{z z'}(r)$ describes the probability of finding a particle of species $z$ at distance $r$ from a particle of species $z'$, 
\begin{eqnarray}\label{eq:rdf}
    g_{z z'}(r) = \frac{{\rm d}N_{z z'}(r)}{4\pi r^2 \rho_{z}\,{\rm d}r}, 
\end{eqnarray}
where $\rho_{z}$ denotes the density of species $z$ and ${\rm d}N_{{z}{z'}}(r)$ is the number of particles of species $z$ found in the spherical shell of radius $r$  and thickness ${\rm d}r$ around a particle of species $z'$.

The ADF is denoted by $P_{zz'z''}(\theta)$, where $\theta$ is the angle determined by relative positions of three particles,  
\begin{eqnarray}\label{eq:adf}
    P_{zz'z''}(\theta; r_{\rm cut}) = \frac{1}{W} \Bigg\< \sum_i\sum_{j\neq i}\sum_{k>j} \delta(\theta -\theta^{{z}{z'}{z''}}_{ijk, r_{\rm cut}})\Bigg\>,
\end{eqnarray}
with the angle
\begin{eqnarray*}
\theta^{{z}{z'}{z''}}_{ijk, r_{\rm cut}} := \cos^{-1} \Bigg(\frac{(
r^{zz'}_{ij})^2+(r^{z'z''}_{jk})^2-(r^{zz''}_{ik})^2}{2r^{zz'}_{ij}r^{z'z''}_{jk}}\Bigg),
\end{eqnarray*}
where $r^{zz'}_{ij}, r^{z'z''}_{jk}, r^{zz''}_{ik}$ represent the distance between two particles associated with species within the cut-off radius $r_{\rm cut}$. In~\eqref{eq:adf}, $W$ normalizes the ADF to ensure it reflects meaningful angular probabilities, independent of system size or particle density. For simplicity, we define $W$ as the total number of particles in the system.

In all numerical experiments conducted in this work, we use {\tt LAMMPS}~\cite{thompson2022lammps} for all-atoms simulations. The coarse-grained simulations were run with {\tt ASE}~\cite{larsen2017atomic} for the implementation of molecular simulation algorithms and an open-source {\tt Julia} package {\tt ACEpotentials.jl}~\cite{witt2023otentials} for the construction of ACE basis and the fitting of ACE-CG models. All the figures of atomistic configurations in this paper are produced using {\tt Ovito}~\cite{stukowski2009visualization}. 
%
The hyperparameters for reproducing the results shown in this paper is collected in the Appendix~\ref{sec:apd:ACE:sub:hyperparameters}.


\subsection{Polymer fluids}
\label{sec:polymer}
In our first numerical example, we simulate star polymer fluids in bulk. Previous studies~\cite{ge2023machine, lei2016data, li2015incorporation} have shown that the coarse-grained (CG) potential $U(\R, \boldsymbol{\zeta})$ for this system exhibits significant many-body interactions, making it unsuitable for accurate representation using only pairwise interactions. While primarily of theoretical interest, this system highlights the strength of ACE-CG in capturing many-body effects.

\subsubsection{Setup}
\label{sec:sub:sub:polymer}

\begin{figure}[htb]
\begin{center}
	\includegraphics[height=6.5cm]{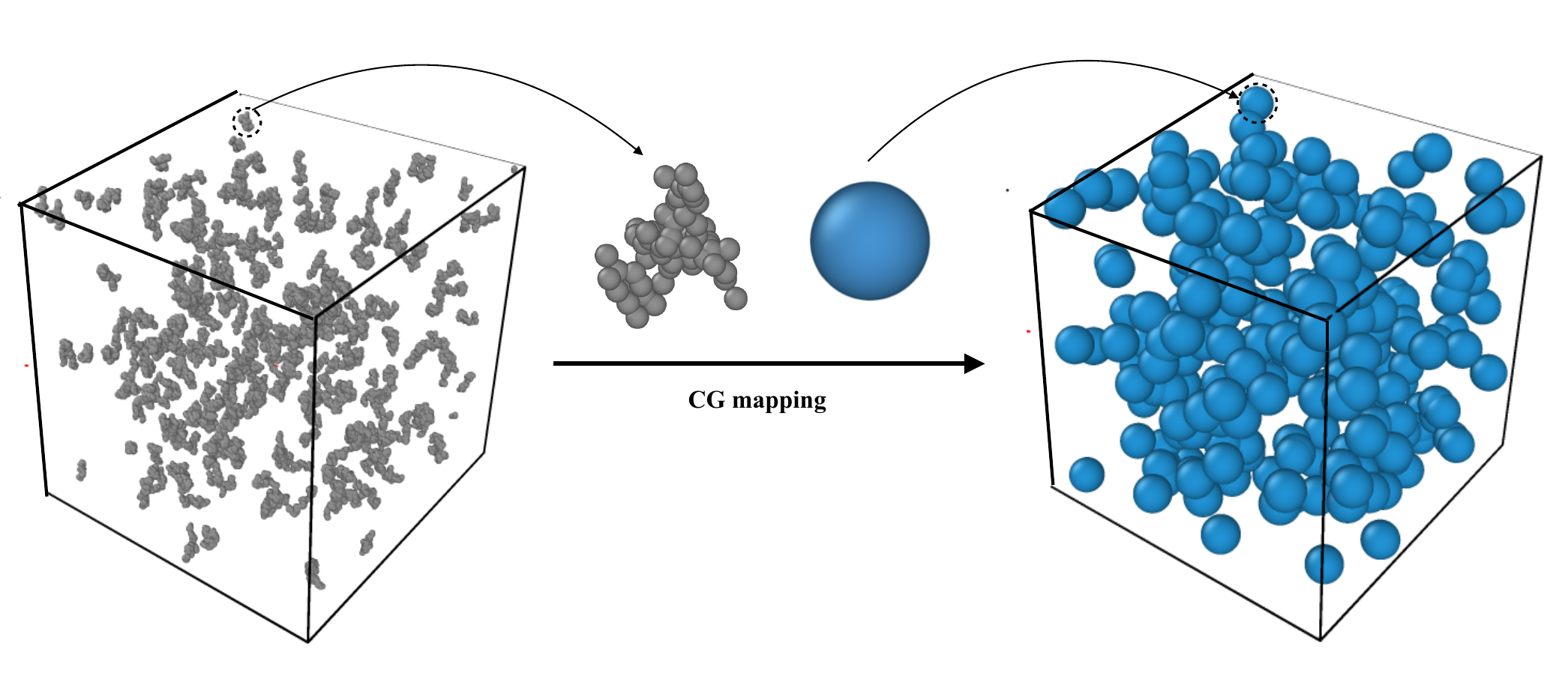}
	\caption{CG mapping for star polymer fluids in bulk.}
	\label{fig:cg-polymer}
\end{center}
\end{figure}

The full system consists of $M$ molecules with a total number of $N$ atoms. Each polymer molecule consists of a ``center" atom connected by $N_{\rm a}$ arms with $N_{\rm b}$ atoms per arm. This polymer molecule is coarse-grained into a single CG particle, as illustrated in Figure~\ref{fig:cg-polymer}. There is only one coarse-grained particle species, which can therefore be ignored.

The potential energy is modeled as a combination of pairwise and bonded interactions, 
\begin{eqnarray}\label{eq:star_poly_V}
    V(\r) = \sum_{i \neq j} V_{\rm p}(r_{ij}) + \sum_{k} V_{\rm b}(l_k),
\end{eqnarray}
where $V_{\rm p}$ is the pairwise interaction between both the intra- and inter-molecular atoms except the bonded pairs and $r_{ij}$ is the distance between the $i$-th and $j$-th atoms. $V_{\rm b}$ is the bonded interaction between neighboring particles of each polymer arm and $l_k$ is the length of the $k$-th bond. The bond potential $V_{\rm b}(l) := \frac{1}{2}k_s(l-l_{0})^2$, where $k_s$ and $l_0$ represents the elastic coefficient and the equilibrium length, respectively.

This relatively simple bead-spring polymer model highlights the critical role of many-body interactions~\cite{li2015incorporation, li2017computing, wang2020data} and the necessity of memory kernels~\cite{lyu2023construction, 2024-friction1} in coarse graining. Thus, it serves as an ideal test system for evaluating our ACE-CG model. In this study, $V_{\rm p}$ takes the form of the Lennard-Jones potential with cut-off $r_{\rm cut}$, i.e.,
\begin{eqnarray}\label{eq:Vp}
	V_{\rm p}(r) =
	\left\{ \begin{array}{ll}
		V_{\rm LJ}(r) - V_{\rm LJ}(r_{\rm cut}), \quad & r < r_{\rm cut}
		\\[1ex]
		0,
		 \quad & r \geq r_{\rm cut}  
	\end{array} \right.,
 \qquad 
 V_{\rm LJ}(r) = 4\varepsilon \left[\Big(\frac{\sigma}{r}\Big)^{12}-\Big(\frac{\sigma}{r}\Big)^6\right],
\end{eqnarray}
where we adopt parameters from~\cite{li2015incorporation}: $N_{\rm a}=12, N_{\rm b}=6, \sigma=2.415$, $\varepsilon=1.0$, $k_s=1.714$, $l_0=2.77$. Also we choose $r_{\rm cut}=2^{1/6}\sigma$ so that $V_{\rm p}$ recovers the Weeks-Chandler-Andersen potential~\cite{ahmed2009phase} (note the entire study is non-dimensional). The full system consists of $N=265$ polymer molecules in a cubic domain with periodic boundary condition imposed along each direction. The Nose-Hoover thermostat is employed to conduct the canonical ensemble simulation with $k_{B}T=3.96$.

\subsubsection{Results}
\label{sec:sub:sub:numerics-polymer}
To evaluate the influence of many-body effects and assess the accuracy of the ACE-CG model, we perform simulations using $U^{\rm ACE}(\bth; \R)$, constructed with different levels of interaction complexity. Specifically, we consider models that include only pairwise interactions by setting $\nu_{\rm max}=1$ (body order 2) in \eqref{eq:acecg_local}, as well as models incorporating higher-order many-body interactions by setting $\nu_{\rm max}=2$ and $\nu_{\rm max}=3$ (respectively, body orders 3 and 4) in \eqref{eq:acecg_local}. This systematic comparison allows us to highlight the importance of many-body terms in accurately capturing the coarse-grained interaction.

\begin{figure}[htb]
\begin{center}
\includegraphics[height=7.0cm]{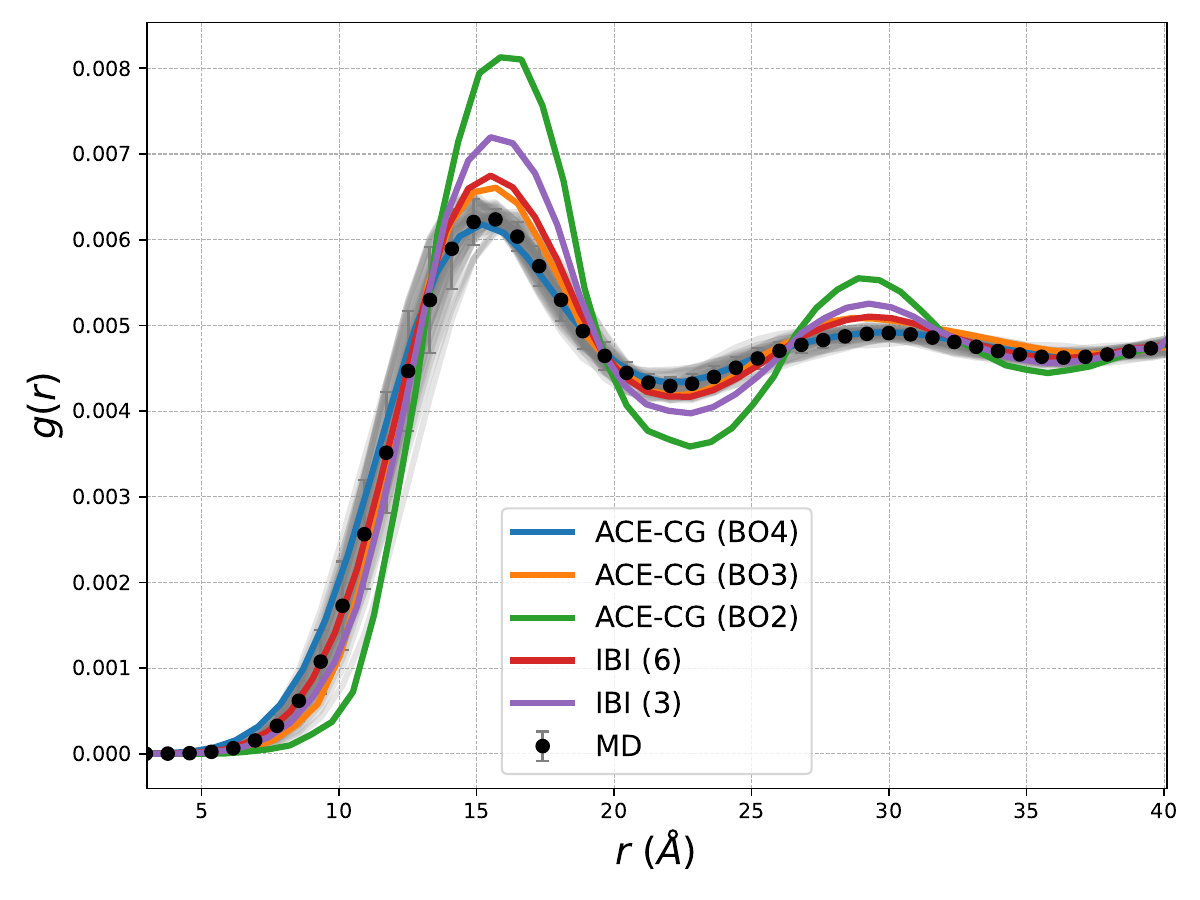}
\caption{Polymer fluids: RDFs generated by different CG models compared to the all-atom MD (black dots).}
	\label{figs:star_poly_RDF}
\end{center}
\end{figure}

We first evaluate the radial distribution functions (RDFs) against the reference all-atom MD results, shown in Figure~\ref{figs:star_poly_RDF}. The reference MD results include error bars and a shaded region, derived from five repeated simulations to account for statistical uncertainty. The ACE-CG model with body order 4 exhibits the best agreement with the mean MD results, while lower body orders show progressively reduced accuracy. Notably, the pairwise CG potential significantly overestimates the RDF peak due to the oversimplification of many-body effects. 

To provide a comparison, we also test the Iterative Boltzmann Inversion (IBI) method~\cite{moore2014derivation}, introduced in Appendix~\ref{sec:apd:sub:IBI}. While the IBI method matches the MD results more closely for RDFs after three (purple line) and six iterations (red line), it does not reach the best order 4 ACE-CG accuracy. Moreover, as we will see below, the accuracy of the IBI method on the simplest many-body observable is very poor.


To quantify the errors in the RDFs, we employed the natural error metric~\cite{pretti2021microcanonical}
\begin{equation}\label{eq:error_RDF}
E_{\rm RDF} := \int_0^{r_{\rm cut}} 4\pi r^2 \big(g(r) - g_{\rm MD}(r) \big)^2 \,{\rm d}r.
\end{equation}
We show the RDF errors of the ACE-CG and IBI models in Table~\ref{tab:rdfDerror-polymer}. We see that increasing the body-order interactions significantly enhances both the qualitative and quantitative agreement with reference data. 

\begin{table}[ht]
    \centering
    \begin{tabular}{|c|c|c|c|c|c|}
        \hline
        CG Models & ACE-CG (BO4) & ACE-CG (BO3) & ACE-CG (BO2) & IBI (6) & IBI (3) \\
        \hline
        $E_{\rm RDF}$ & 0.058 & 0.152 & 4.711 & 0.183 & 0.845 \\
        \hline
    \end{tabular}
    \vskip0.3cm
    \caption{RDF errors for star-polymer systems. The errors quantify deviations from the reference MD data.}
    \label{tab:rdfDerror-polymer}
\end{table}

\begin{figure}[htb]
\centering
\begin{subfigure}[b]{0.48\textwidth}
    \centering
    \includegraphics[height=6.0cm]{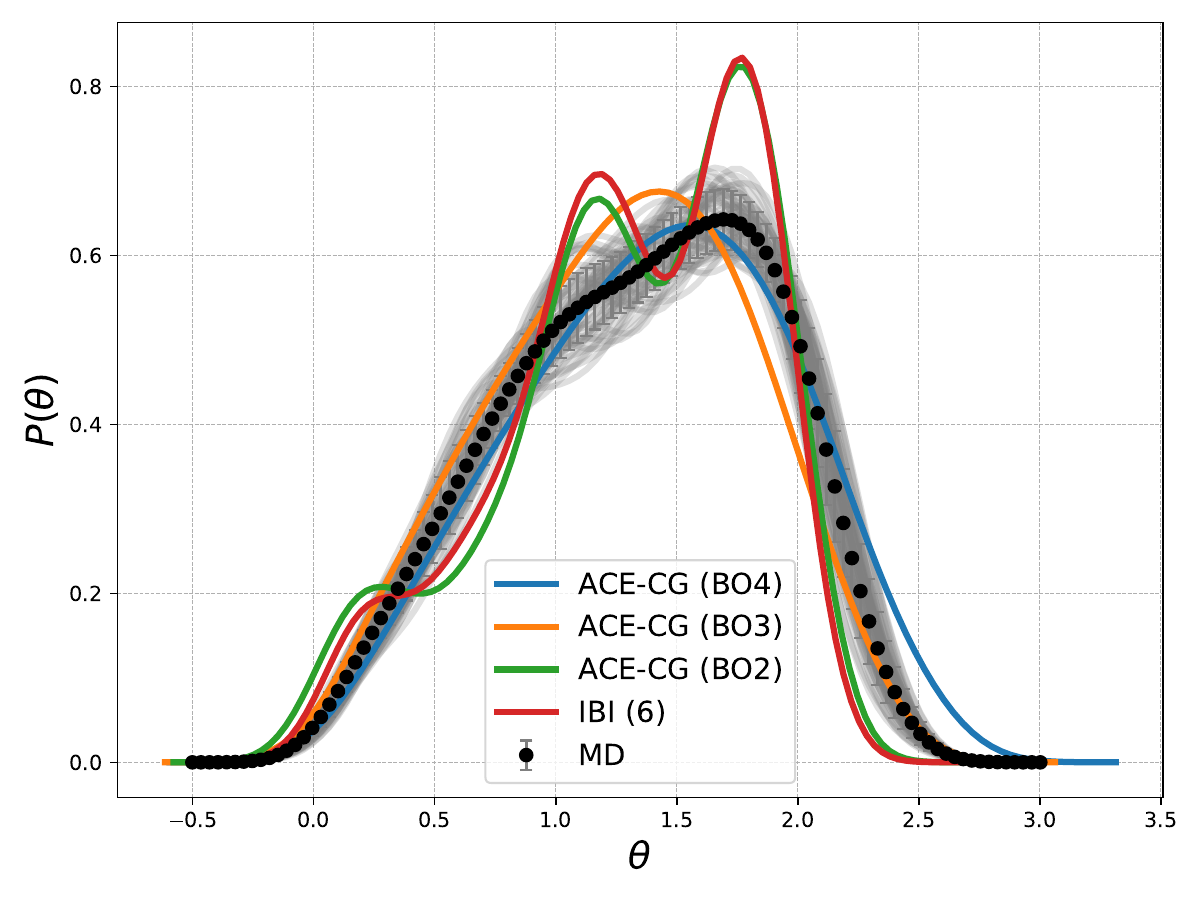}
    \caption{$r_{\rm cut}=15\,\mathring{\textrm{A}}$.}
    \label{fig:star_poly_ADF_13}
\end{subfigure}
\hfill
\begin{subfigure}[b]{0.48\textwidth}
    \centering
    \includegraphics[height=6.0cm]{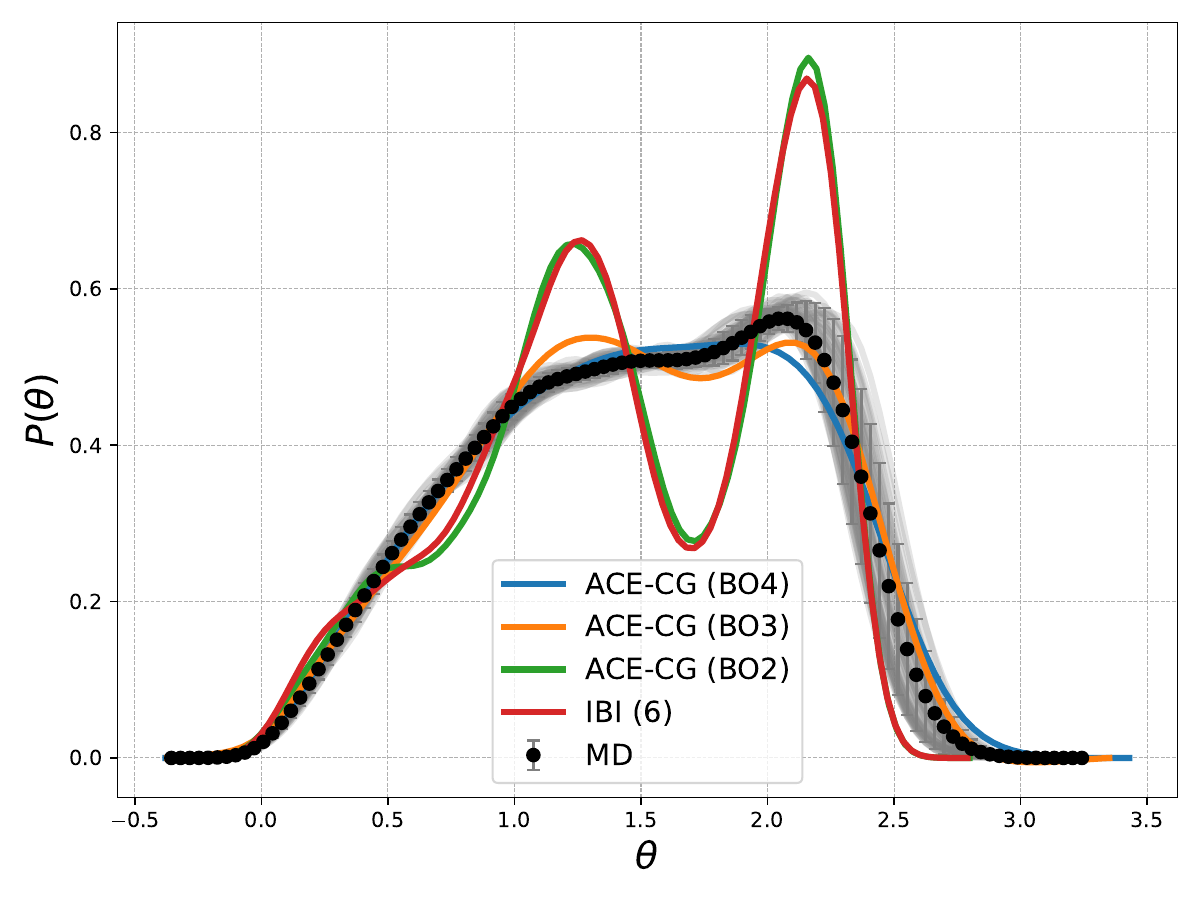}
    \caption{$r_{\rm cut}=22\,\mathring{\textrm{A}}$.}
    \label{fig:star_poly_ADF_20}
\end{subfigure}
\caption{Polymer fluids: ADFs for different cutoff radii.}
\label{figs:star_poly_ADF}
\end{figure}

Next, we turn to the error in predicting the angular distribution functions (ADFs). Figure~\ref{figs:star_poly_ADF} shows the ADF results for two cutoff values, $r_{\rm cut}=15$ and $r_{\rm cut}=22$, with additional results for $r_{\rm cut}=30$ provided in the appendix (cf.~Figure~\ref{figs:star_poly_ADF_appendix}). These cutoff values were chosen to align with the peaks and troughs of the RDF shown in Figure~\ref{figs:star_poly_RDF}. Similar to the RDF analysis, the ACE-CG model with body order 4 demonstrates the best agreement with the full MD results. The decreasing of body order results in progressively larger deviations from the full MD results. The IBI method, which is limited to pair interactions, performs poorly for ADF predictions. 

To quantify the error in the angular distribution function (ADF), we use the following metric:
\begin{equation}\label{eq:error_ADF}
E_{\text{ADF}} := \int_{0}^{\pi} \sin(\theta) \big( P(\theta) - P_{\text{MD}}(\theta) \big)^2\,{\rm d}\theta.
\end{equation}
Table~\ref{tab:adf_error-polymer} reports the ADF errors for three different cutoff values. The results further emphasize the importance of higher body orders in accurately modeling the CG representations of this star-polymer system. They also demonstrate the ability of the ACE-CG model to systematically and accurately capture those many-body interactions with increasing model complexity. 

\begin{table}[ht]
    \centering
    \begin{tabular}{|c|c|c|c|c|}
        \hline
        CG Models & ACE-CG (BO4) & ACE-CG (BO3) & ACE-CG (BO2) & IBI (6) \\
        \hline
        $E_{\rm ADF}$, $r_{\rm cut}=15$ & 0.095 & 0.282 & 0.420 & 0.458 \\
        \hline
        $E_{\rm ADF}$, $r_{\rm cut}=22$ & 0.052 & 0.109 & 0.682 & 0.683 \\
        \hline
        $E_{\rm ADF}$, $r_{\rm cut}=30$ & 0.018 & 0.025 & 0.246 & 0.250 \\
        \hline
    \end{tabular}
    \vskip0.3cm
    \caption{\label{tab:adf_error-polymer} ADF errors for star-polymer systems. The errors quantify deviations from the reference MD data.}
\end{table}

Finally, we evaluate the computational efficiency of the ACE-CG models. Table~\ref{tab:star_poly} compares the simulation speeds of the all-atom model and the ACE-CG models with different body orders. The results reveal a significant improvement in simulation speed when using the ACE-CG models, emphasizing its ability to balance computational efficiency and accuracy effectively. This efficiency is expected to become even more pronounced in simulations of more complex polymer interactions and realistic polymer systems, further showcasing the potential of the ACE-CG approach for large-scale applications.

\begin{table}[htbp]
\centering
\begin{tabular}{|c|c|c|c|c|}
\hline
Models & time step (fs) & $N_{\rm site}$ & ms/step & ns/day \\ \hline
MD & 1 & 19345 & 66.5 & 1.3 \\ 
ACE-CG (BO2) & 10 & 265 & 7.9 (2.1) & 108.5 (410) \\ 
ACE-CG (BO3) & 10 & 265 & 11.3 (2.7) & 76.4 (323) \\ 
ACE-CG (BO4) & 10 & 265 & 14.1 (4.5) & 61.2 (190) \\ 
\hline
\end{tabular}
\vskip0.2cm
\caption{\label{tab:star_poly} Simulation speeds for polymer fluids. Numbers in brackets are simulation times with optimized computational kernels that are not yet available through ASE or LAMMPS at the time of writing.}
%
%
\end{table} 

%


\subsection{Methanol fluids}
\label{sec:sub:methanol}
Next, we investigate methanol fluids, the simplest form of alcohol consisting of a methyl group (CH3) and a hydroxyl group (OH). Although previous studies, such as those by Noid et al.~\cite{noid2013perspective, noid2008multiscale}, often used CG models with a single-site representation for each molecule, we adopt a higher resolution approach. Specifically, we employ a two-site representation, placing one site at the center of mass of the methyl group and the other at the center of mass of the hydroxyl group. This enhanced resolution allows us to better capture molecular interactions and evaluate the impact of different levels of coarse-graining on reproducing all-atom system properties. 

\subsubsection{Setup}
\label{sec:sub:sub:methanol_system}

\begin{figure}[htb]
\begin{center}
\includegraphics[height=5.5cm]{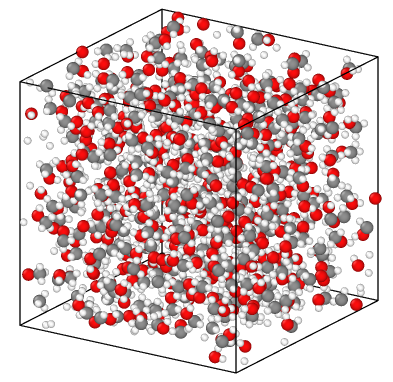} \quad 
	\includegraphics[height=4.75cm]{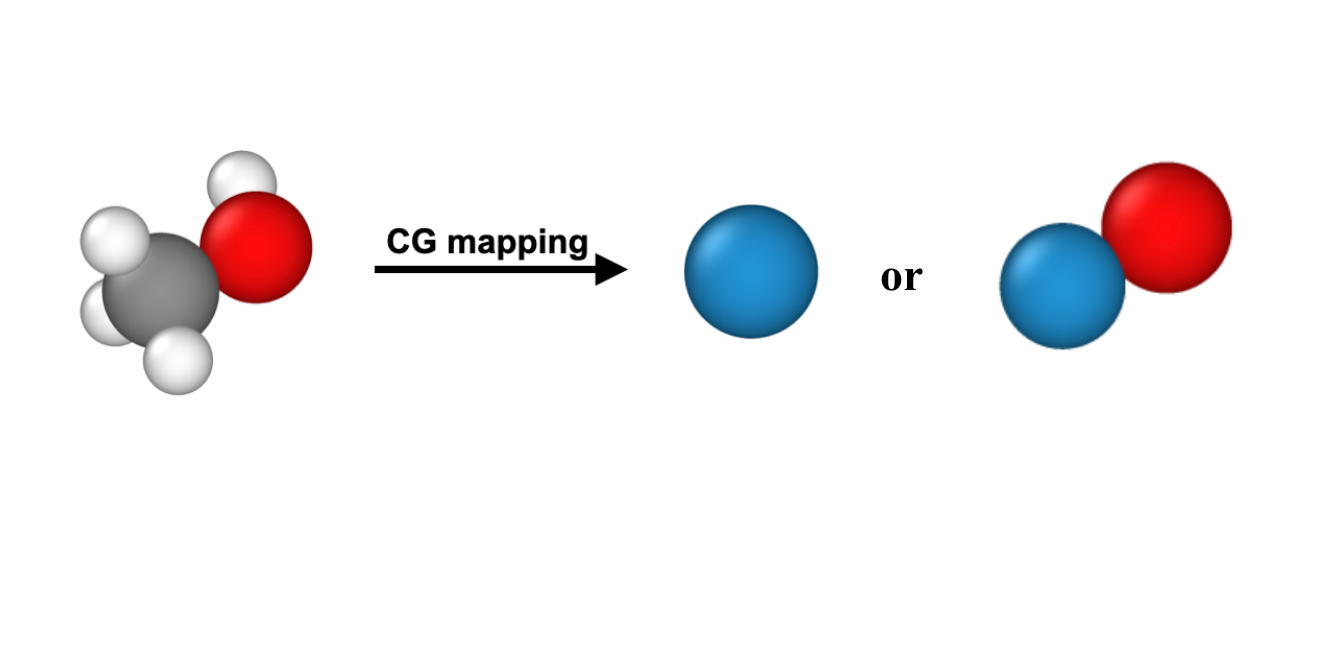}
	\caption{Structure of bulk methanol fluids. Left: the snapshot of MD trajectory at $t=1.0$ ns; Right: two levels of CG mappings for methanol.}
	\label{figs:methanol}
\end{center}
\end{figure}

We investigate the size effects in training CG models by employing a training domain that is smaller than the testing domain, but has the same density. This generalization capability is essential for machine-learned CG force fields, enabling their application to larger and more complex systems.
We perform all-atom MD simulations on a small training domain consisting of $N=50$ methanol molecules in a cubic periodic cell of $15\mathring{\rm A} \times 15\mathring{\rm A} \times 15\mathring{\rm A}$. The testing system doubles the size of the training domain to ($30\mathring{\rm A} \times 30\mathring{\rm A} \times 30\mathring{\rm A}$), containing $N=400$ methanol molecules while maintaining the same density and interaction parameters.
Simulations are carried out at $T = 300$~K using the {\tt OPLS-AA} force field~\cite{kaminski2001evaluation}. A time step of $1.0$~fs is employed to ensure accurate dynamics. The simulation protocol includes: (i) energy minimization using the conjugate gradient algorithm until convergence; (ii) equilibration in the NPT ensemble for $1e5$ steps; and (iii) production runs in the NVT ensemble for $2e6$ steps. A snapshot of the MD trajectory at $t=1.0$~ns is shown in Figure~\ref{figs:methanol}.

\subsubsection{Results}
\label{sec:sub:sub:numerics_methanol}

The ACE-CG potential ${U}^{\rm ACE}(\bth; \R, \boldsymbol{\zeta})$ is trained using the approach outlined in Section~\ref{sec:sub:sub:params}, employing training data acquired from MD simulations. Consistent with the analysis in Section~\ref{sec:polymer}, we evaluated ACE-CG models with varying body orders to investigate the many-body effects in this more realistic methanol fluid system.

\begin{figure}[htb]
\begin{center}
\includegraphics[height=7.0cm]{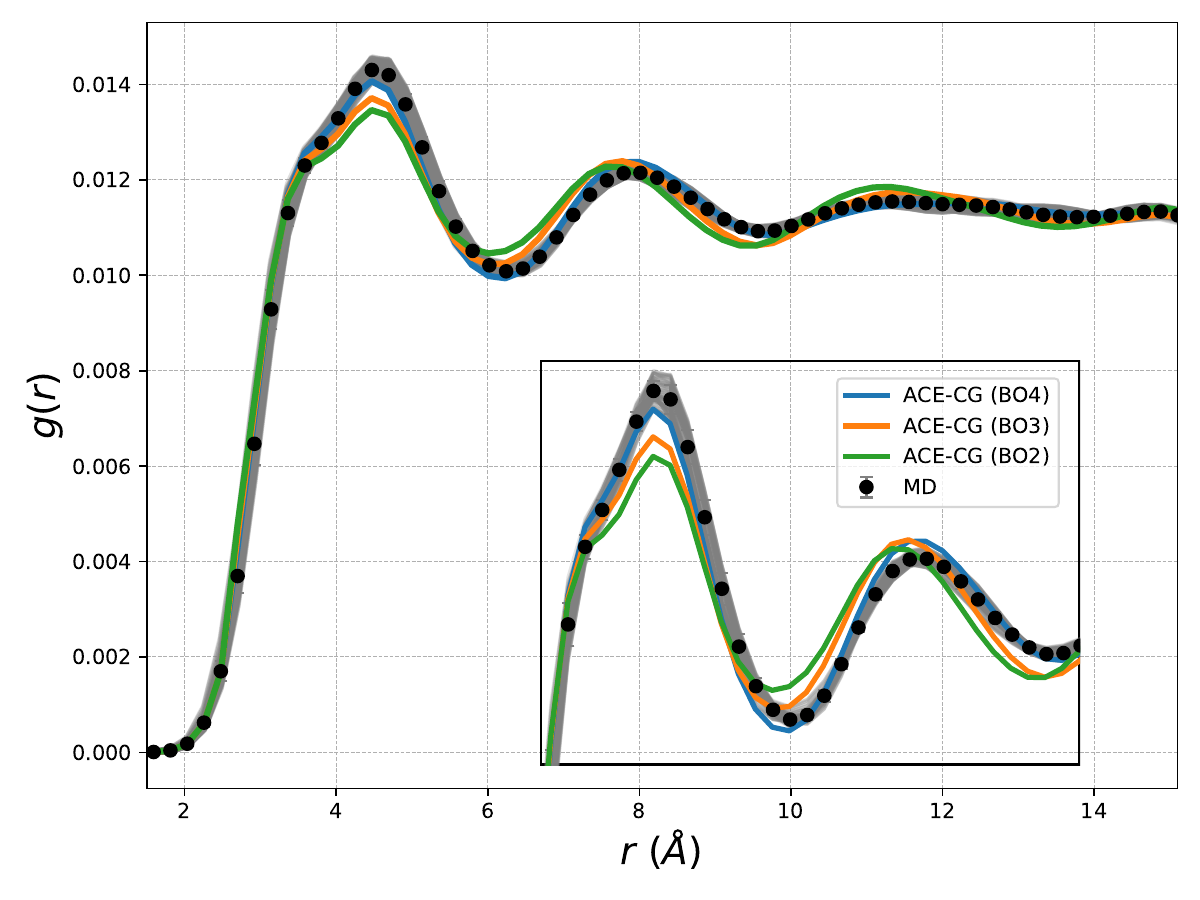}
\caption{Methanol fluids: RDF for one-site description.}
	\label{figs:pure_methanol_RDF_1}
\end{center}
\end{figure}

\begin{figure}[htb]
    \centering
    \begin{subfigure}[t]{0.48\textwidth}
        \centering
        \includegraphics[height=5.5cm]{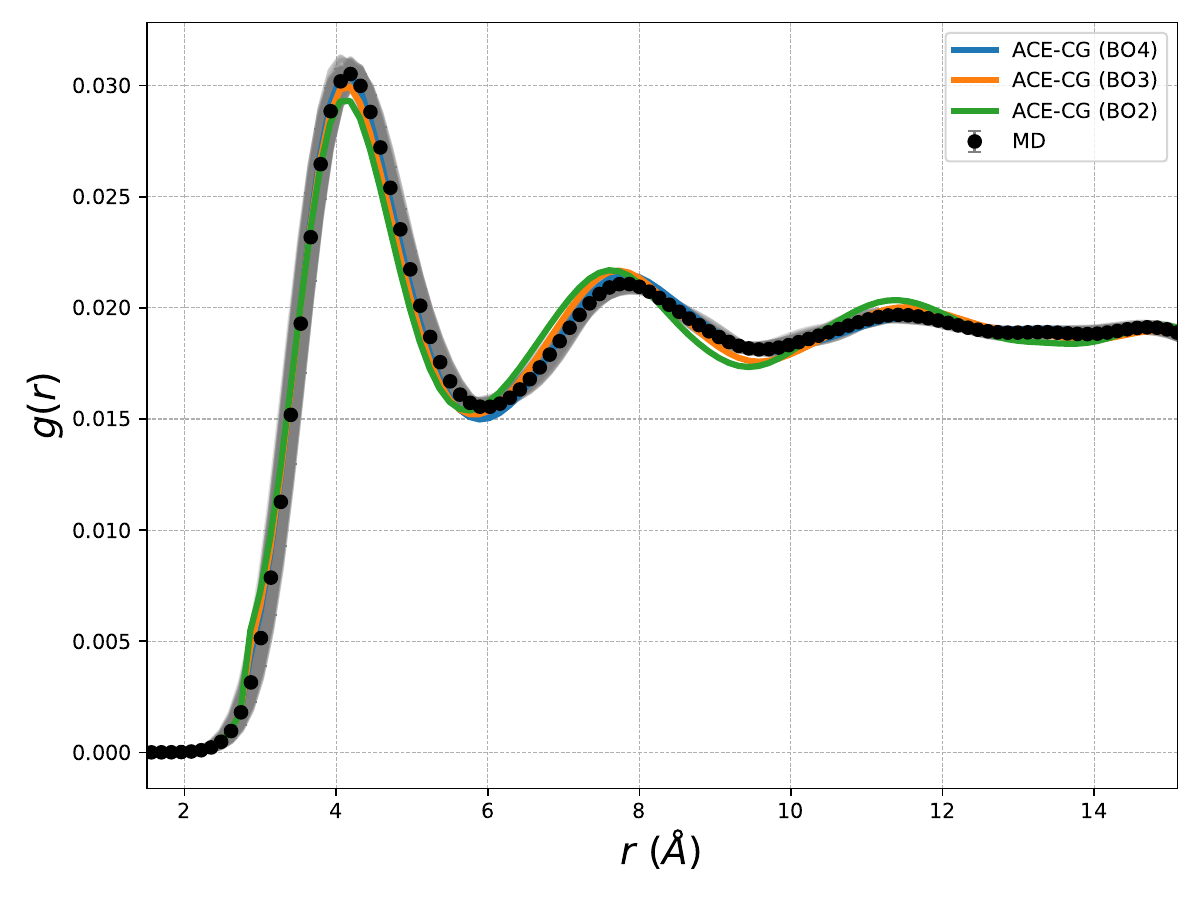}
        \caption{Carbon-carbon RDF.}
        \label{figs:pure_methanol_RDF_2_C}
    \end{subfigure}
    \hfill
    \begin{subfigure}[t]{0.48\textwidth}
        \centering
        \includegraphics[height=5.5cm]{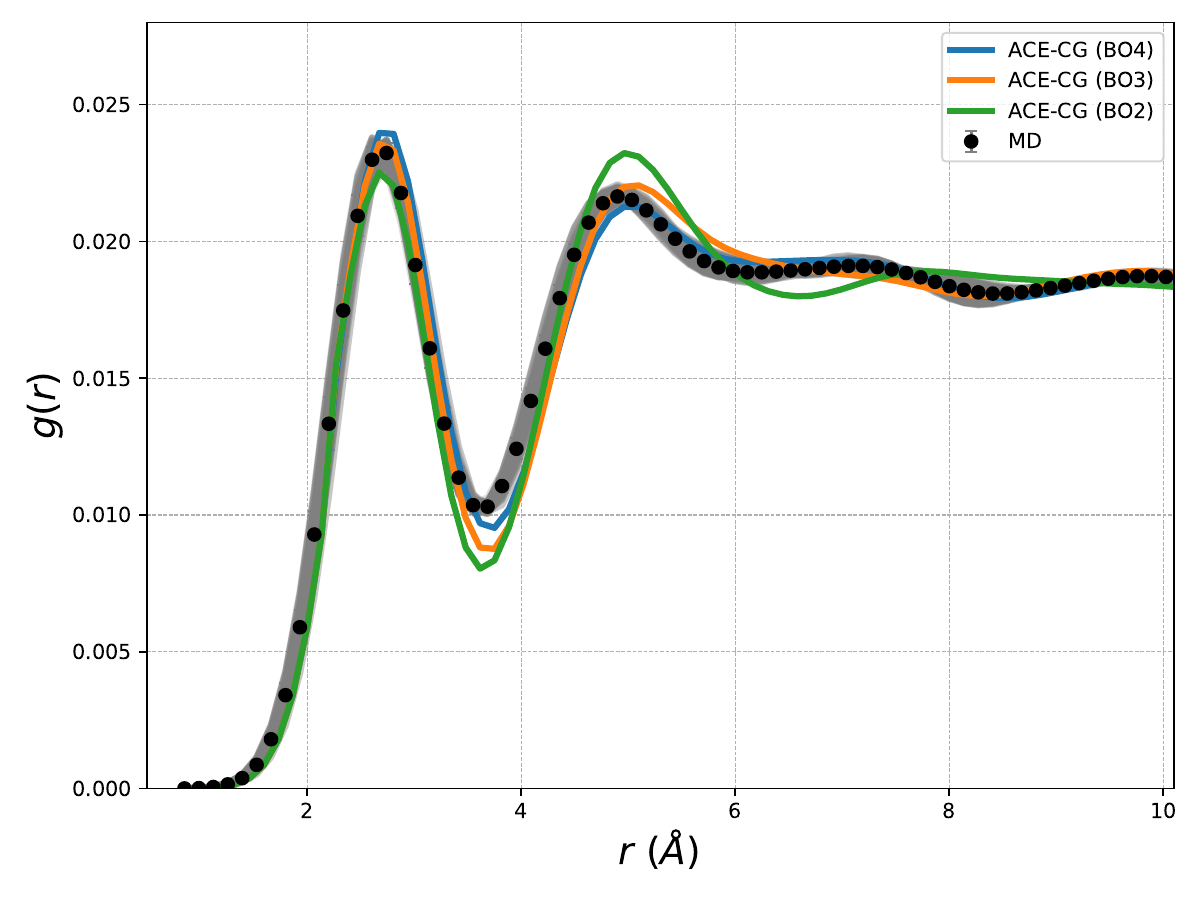}
        \caption{Oxygen-oxygen RDF.}
        \label{figs:pure_methanol_RDF_2_O}
    \end{subfigure}
    \caption{Methanol fluids: RDFs for the two-site description.}
    \label{figs:pure_methanol_RDF_2}
\end{figure}

Figures~\ref{figs:pure_methanol_RDF_1} and \ref{figs:pure_methanol_RDF_2} show the RDFs computed using \eqref{eq:rdf}, highlighting the effects of different CG representations. Figure~\ref{figs:pure_methanol_RDF_1} corresponds to a one-site CG mapping, while Figure~\ref{figs:pure_methanol_RDF_2} illustrates a two-site mapping, including RDFs for carbon-carbon and oxygen-oxygen groups. Mixed carbon-oxygen RDFs are provided in the supplement (cf.~Figure~\ref{figs:pure_methanol_RDF_2_appendix}). Similar to the polymer case, RDF errors are evaluated using the metric defined in~\eqref{eq:error_RDF}, with the results summarized in Table~\ref{tab:rdfDerror}. The values in parentheses represent size-consistent results, where the training and testing systems are identical in size, providing a direct evaluation of the generalization accuracy of the model. In both CG representations, we evaluated three ACE-CG models: ACE-CG (BO2), which includes only pairwise interactions (body order two), and ACE-CG (BO3) and ACE-CG (BO4), which incorporate higher-order many-body interactions. These RDFs were benchmarked against reference MD simulations, providing a comprehensive assessment of the accuracy and performance of the ACE-CG models.

We derive three key insights from Figure~\ref{figs:pure_methanol_RDF_1}, Figure~\ref{figs:pure_methanol_RDF_2}, and Table~\ref{tab:rdfDerror}. First, the generalization approach performs well, as evidenced by the moderate changes observed between size-consistent and size-extensive results, with only a negligible reduction in accuracy. 
Second, the two-site representation consistently outperforms the one-site approach, as clearly and quantitatively shown in Table~\ref{tab:rdfDerror}. This result is expected since higher-resolution representations offer more degrees of freedom to capture the fine-grained details of the all-atom system. Third, across both CG mappings, the ACE-CG models demonstrate strong agreement with the reference RDFs, with ACE-CG (BO4) providing the highest accuracy. Notably, unlike the polymer case where many-body effects are pronounced, the methanol results indicate that pairwise interactions alone yield qualitatively reasonable outcomes. This suggests that many-body effects are less critical to capture the RDF for this system. However, as Table~\ref{tab:rdfDerror} shows, including higher body orders in the CG model still leads to clear improvements in quantitative accuracy. 

\begin{table}[htbp]
\centering
\begin{tabular}{|c|c|c|c|}
\hline
Model & ACE-CG (BO4) & ACE-CG (BO3) & ACE-CG (BO2) \\ \hline
$E_{\rm RDF}$, one-site & 0.009 (0.008) & 0.053 (0.045) & 0.165 (0.152) \\
\hline
$E_{\rm RDF}$, two-sites & 0.005 (0.003) & 0.021 (0.016) & 0.064 (0.058) \\ 
\hline
\end{tabular}
\vskip0.2cm
\caption{\label{tab:rdfDerror} RDF errors for methanol fluids across different ACE-CG models. The values in parentheses denote size-consistent results, where the training and testing systems are of identical size, providing a direct measure of generalization.}
\end{table}

\begin{figure}[htb]
    \centering
    \begin{subfigure}[b]{0.32\textwidth}
        \centering
        \includegraphics[height=4.1cm]{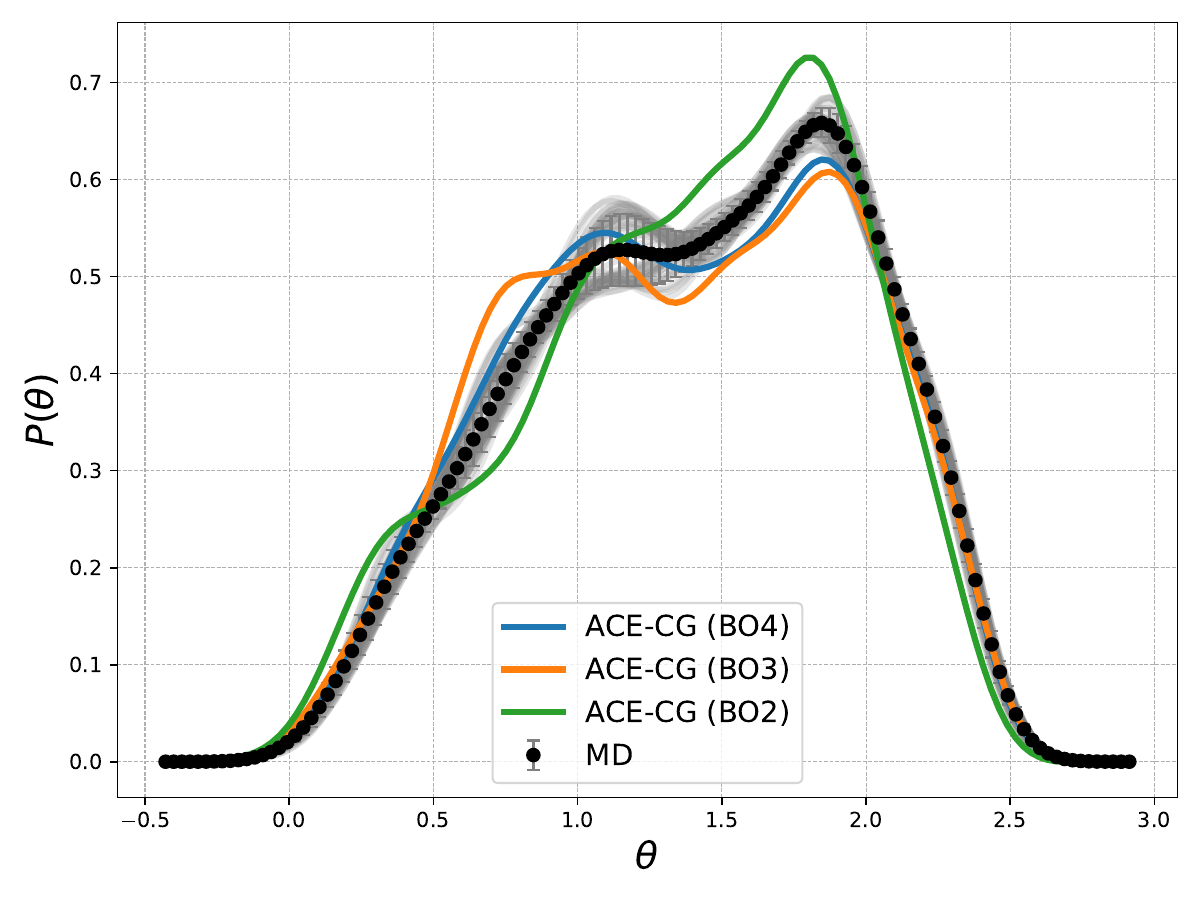}
        \caption{$r_{\rm cut} = 4.5 \mathring{\textrm{A}}$}
        \label{fig:adf_local_1_4.5}
    \end{subfigure}
    \begin{subfigure}[b]{0.32\textwidth}
        \centering
        \includegraphics[height=4.1cm]{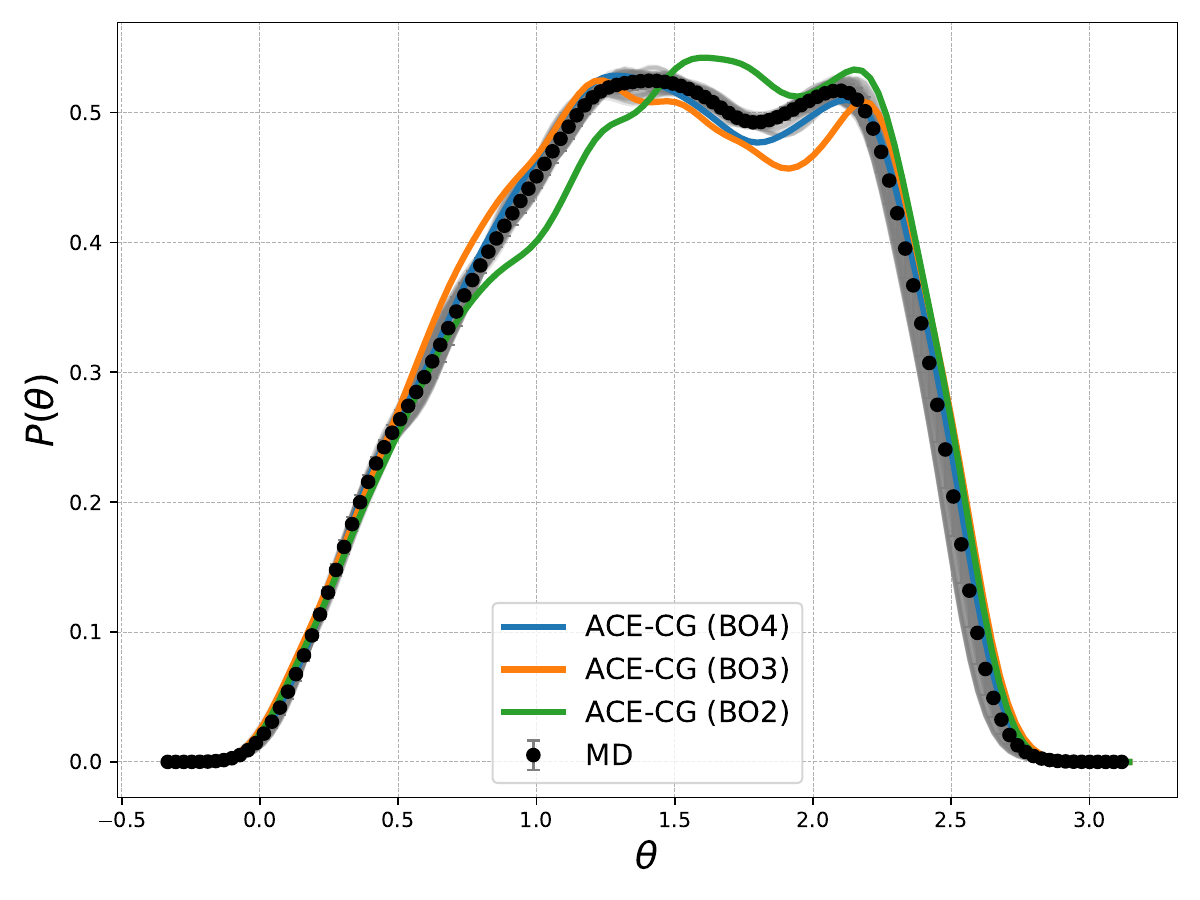}
        \caption{$r_{\rm cut} = 6.0 \mathring{\textrm{A}}$}
        \label{fig:adf_local_1_6.0}
    \end{subfigure}
    \begin{subfigure}[b]{0.32\textwidth}
        \centering
        \includegraphics[height=4.1cm]{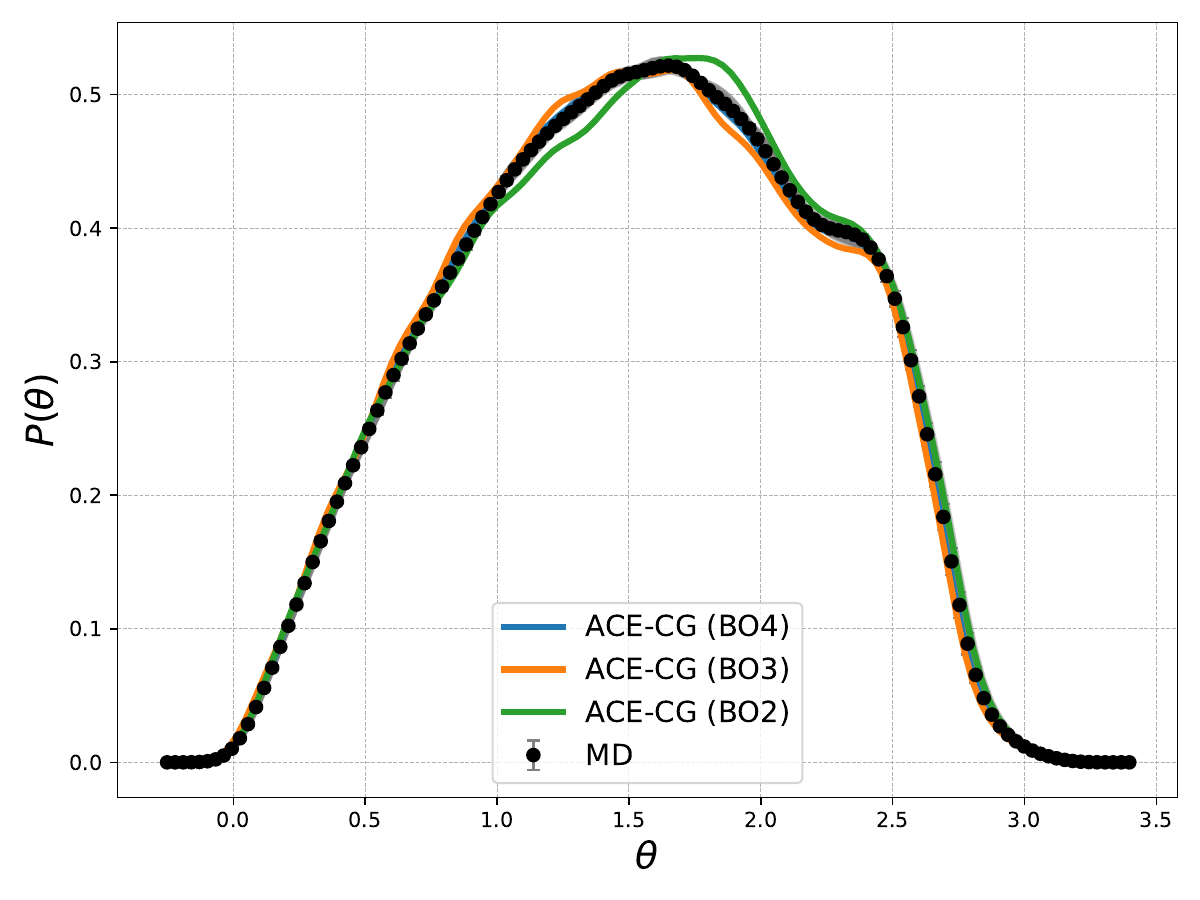}
        \caption{$r_{\rm cut} = 8.0 \mathring{\textrm{A}}$}
        \label{fig:adf_local_1_8.0}
    \end{subfigure}
    \caption{Methanol fluids: one-site ADF for different cutoff radii.}
    \label{figs:pure_methanol_ADF_local}
\end{figure}

\begin{figure}[htb]
    \centering
    \begin{subfigure}[b]{0.32\textwidth}
        \centering
        \includegraphics[height=4.1cm]{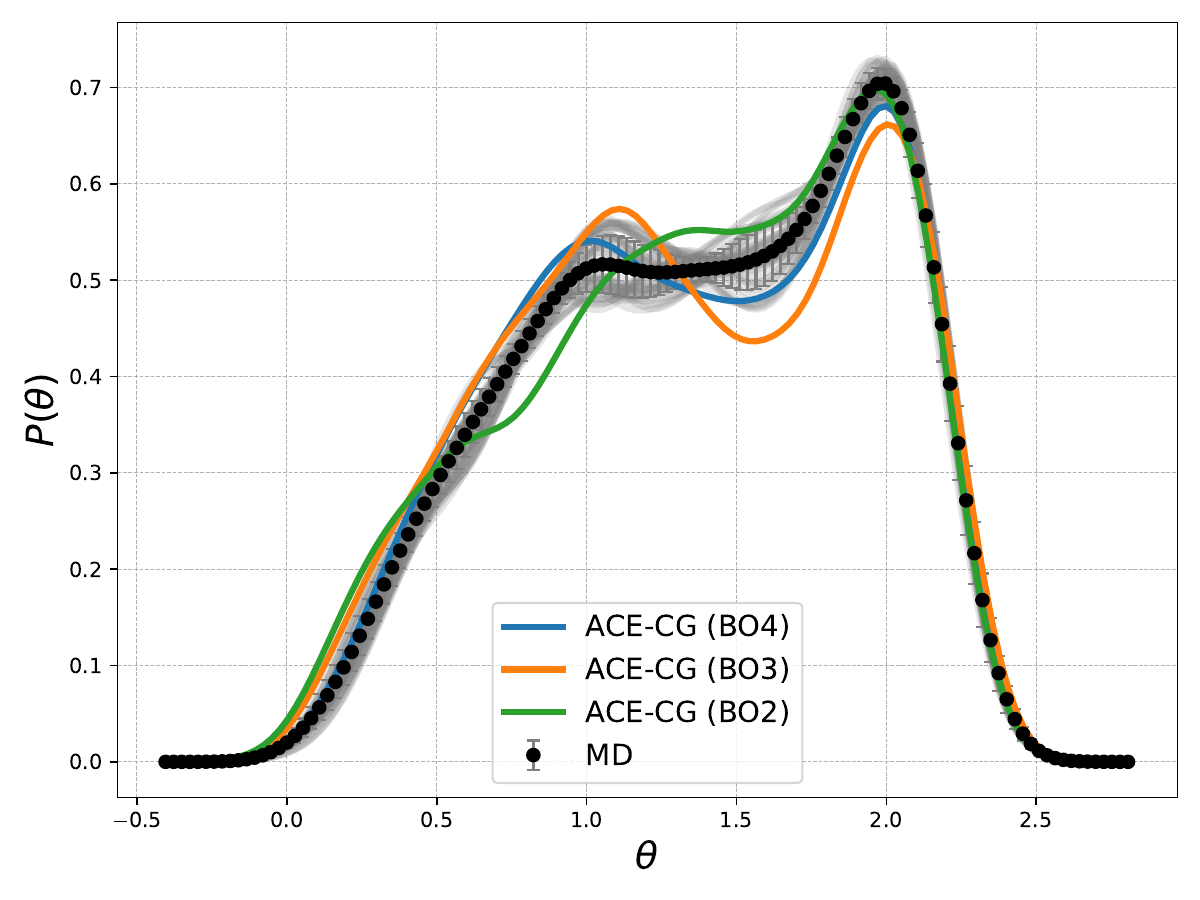}
        \caption{$r_{\rm cut} = 4.5 \mathring{\textrm{A}}$}
        \label{fig:adf_local_2_4.5_C}
    \end{subfigure}
    \begin{subfigure}[b]{0.32\textwidth}
        \centering
        \includegraphics[height=4.1cm]{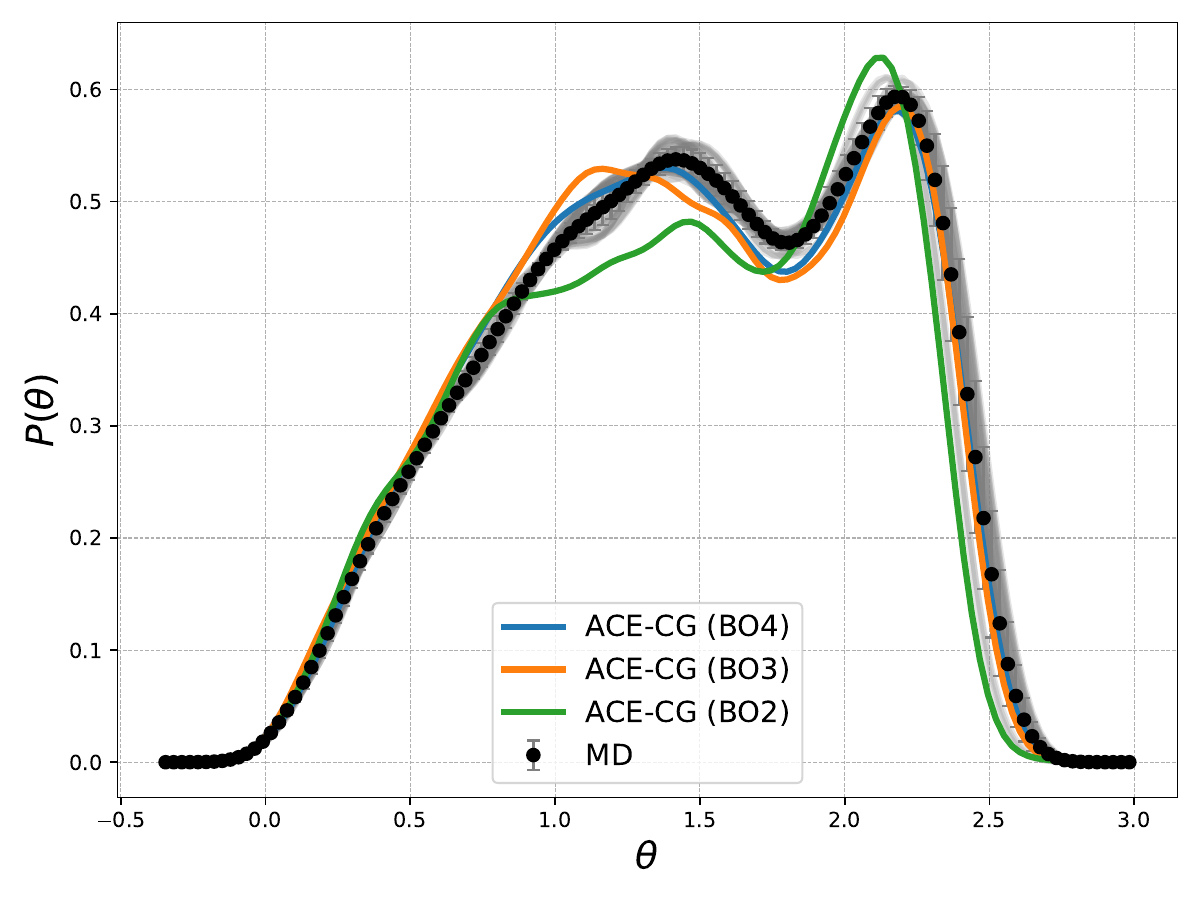}
        \caption{$r_{\rm cut} = 6.0 \mathring{\textrm{A}}$}
        \label{fig:adf_local_2_6.0_C}
    \end{subfigure}
    \begin{subfigure}[b]{0.32\textwidth}
        \centering
        \includegraphics[height=4.1cm]{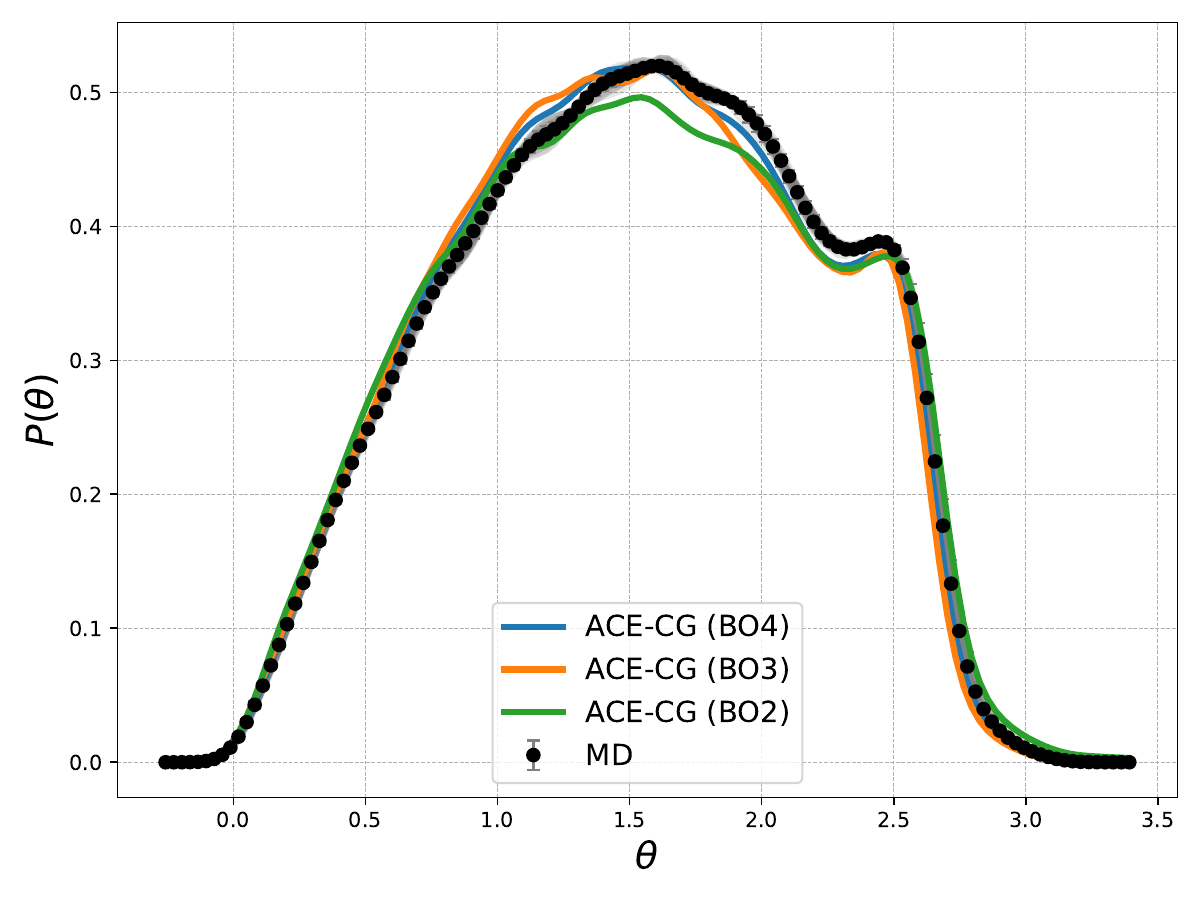}
        \caption{$r_{\rm cut} = 8.0 \mathring{\textrm{A}}$}
        \label{fig:adf_local_2_8.0_C}
    \end{subfigure}
    \caption{Methanol fluids: two-sites ADF for carbon group with different cutoff radii.}
    \label{figs:pure_methanol_ADF_2_local}
\end{figure}

Figure~\ref{figs:pure_methanol_ADF_local} and Figure~\ref{figs:pure_methanol_ADF_2_local} present the angular distribution functions (ADFs) calculated using three cutoff values, $r_{\rm cut}=4.5, 6.0, 8.0\mathring{\textrm{A}}$. For clarity, we only show the carbon group ADFs in the main text, leaving the oxygen group ADFs to the supplement (cf.~Figure~\ref{figs:pure_methanol_ADF_2_OOO_appendix}). To ensure statistical reliability, the all-atom MD results are shown with error bars derived from 5 repeated simulations. Additionally, we quantify the ADF deviations using the error metric defined in~\eqref{eq:error_ADF}, with the results summarized in Table~\ref{tab:adfDerror}. Parenthetical values represent size-consistent results, where the training and testing systems have the same size, providing a benchmark for evaluating generalization accuracy.
The body order 2 and 3 ACE-CG model predictions of the ADF still align with the reference MD results, but much more crudely than for the RDF. By contrast the body order 4 model captures the ADF very well. This is quantified more clearly in Table~\ref{tab:adfDerror}, which shows a marked drop in the approximation error when the body order is increased to four. 
%

Comparing the one-site and two-site CG mappings also reveals variations in the model’s accuracy based on the chosen CG representation, as shown quantitatively in Table~\ref{tab:adfDerror}. This demonstrates the critical role of selecting an appropriate CG mapping level to balance computational efficiency and accuracy. 

\begin{table}[htbp]
\centering
\begin{tabular}{|c|c|c|c|}
\hline
Model & ACE-CG (BO4) & ACE-CG (BO3) & ACE-CG (BO2) \\ \hline
$E_{\rm ADF}$ (one-site, $r_{\rm cut}=4.5\mathring{\textrm{A}}$) & 0.091 (0.083) & 0.273 (0.258) & 0.298 (0.272) \\
$E_{\rm ADF}$ (one-site, $r_{\rm cut}=6.0\mathring{\textrm{A}}$) & 0.050 (0.044) & 0.139 (0.129) & 0.168 (0.150) \\
$E_{\rm ADF}$ (one-site, $r_{\rm cut}=8.0\mathring{\textrm{A}}$) & 0.019 (0.016) & 0.071 (0.065) & 0.087 (0.075) \\
\hline 
$E_{\rm ADF}$ (two-sites, $r_{\rm cut}=4.5\mathring{\textrm{A}}$) & 0.148 (0.143) & 0.214 (0.202) & 0.178 (0.158) \\ 
$E_{\rm ADF}$ (two-sites, $r_{\rm cut}=6.0\mathring{\textrm{A}}$) & 0.047 (0.042) & 0.113 (0.105) & 0.154 (0.141) \\ 
$E_{\rm ADF}$ (two-sites, $r_{\rm cut}=8.0\mathring{\textrm{A}}$) & 0.010 (0.009) & 0.059 (0.056) & 0.072 (0.068) \\ 
\hline
\end{tabular}
\vskip0.2cm
\caption{\label{tab:adfDerror} ADF errors for methanol fluids across different ACE-CG models and cutoff distances. Parenthetical values represent size-consistent results, where the training and testing systems have the same size, providing a benchmark for evaluating generalization accuracy.}
\end{table}

Finally, Table~\ref{tab:methanol} compares the computational performance of all-atom MD and the ACE-CG (BO4) model, expressed in terms of simulated nanoseconds per day. The ACE-CG model again demonstrates a significant improvement in computational efficiency compared to all-atom simulations, offering faster simulation speeds while maintaining high accuracy in reproducing the structural distributions observed in the reference system. These results underscore the capability of the ACE-CG approach to achieve a practical balance between computational speed and fidelity to all-atom behavior. 

\begin{table}[htbp]
\centering
\begin{tabular}{|c|c|c|c|c|}
\hline
Model & time step (fs) & $N_{\rm site}$ & ms/step & ns/day \\ \hline
MD & 1 & 2400 & 20.5 & 4.2 \\ 
one-site & 4 & 400 & 6.8 (1.3) & 50.8 (262) \\ 
two-sites & 4 & 800 & 18.4 (3.5) & 18.7 (69.2) \\
\hline
\end{tabular}
\vskip0.2cm
%
%
\caption{\label{tab:methanol} Computational performance for methanol fluids, where the results of one-site and two-sites ACE-CG (BO4) model are compared against all atom molecular dynamics simulations. Numbers in brackets indicate improved simulation times using optimized computational kernels that are not yet available through ASE or LAMMPS at the time of writing.}
\end{table}

\section{Conclusion}
\label{sec:conclusion}
We introduced ACE-CG for developing many-body coarse-grained models based on the Atomic Cluster Expansion (ACE) framework, a widely used technique in machine-learned interatomic potentials (MLIPs). 
The method achieves a relatively low-cost but accurate and systematically improvable and size extensive representation of many-body interactions, enabling accurate modeling of equilibrium properties within coarse-grained molecular dynamics simulations. Our results demonstrate 
the significance of many-body effects in CG systems for the accurate predictions of equilibrium properties, while also highlighting the efficiency of our approach.

The ACE framework offers a promising avenue for enhancing the computational feasibility of parameterizing CGMD models, while retaining essential information about their complex many-body interactions. The ACE framework is in principle general and flexible enough to allow a large number of possible extensions such as the addition of memory kernels (see \cite{2024-friction1} for a step in this direction) or incorporating more detailed geometric information about coarse-grained particles (anisotropy). 








\appendix

\section{Theoretical Background}
\label{sec:apd:preliminaries}
\renewcommand{\theequation}{A.\arabic{equation}}

In this section, we provide an overview of the concepts and formulations relevant to Section~\ref{sec:method} and address aspects that have not been covered in the main text. These include the definition of a consistent coarse-grained (CG) model, the derivation of the mean force and instantaneous collective force, and a brief introduction to the iterative Boltzmann inversion method.

\subsection{Consistent CG model}

First of all, we provide an overview of the concepts and formulations that are relevant to Section~\ref{sec:method}. The foundation of our approach draws inspiration from previous works~\cite{john2017many, wang2019machine, zhang2018deepcg}, which have explored different types of MLIPs in the context of CG.

\subsubsection{CG mapping}
\label{appendix:sub:sub:cg_mapping}

We assume an explicit connection between the atomistic and corresponding CG models through projections $\{\mathcal{M}_1, \ldots, \mathcal{M}_N\}$ such that $\R_I = \mathcal{M}_I(\r)$. 
The projections are many-to-one mappings: many atomistic configurations are mapped to the same CG configuration. For the sake of simplicity of presentation we assume that the projections are linear in the position, 
$\R_I = \mathcal{M}_I(\r) 
        = \sum_{i \in \mathcal{J}_I} w_{Ii}  \r_i,$
with $\sum_{i} w_{Ii} = 1$, where $\mathcal{J}_I$ denotes the set of atoms that form the CG particle $I$. The center of mass projection is particularly prevalent throughout the literature; it can be expressed as 
\begin{equation} \label{eq:ctr_mass_appendix}    
    w_{Ii} = \begin{cases}
        \frac{m_i}{\sum_{j \in \mathcal{J}_I} m_j}, & \text{if } i \in \mathcal{I}_I, \\ 
        0, & \text{otherwise}. 
    \end{cases} 
\end{equation}
Much more general choices are possible~\cite{voth2008coarse}.

For the momentum we have $\P_I = \mathcal{M}_{I}(\p)$. Coarse-grained position and momentum are connected by $\P_I = M_I \cdot {\rm d}\R_I/{\rm d}t$, where $M_I$ is the {\it effective mass} of CG site $I$. The linear mapping induces the following relation for the CG momenta:
\begin{equation*}
    \mathcal{M}_{I}(\p^n) = M_I \sum_{i=1}^{n} w_{Ii} \frac{\p_i}{m_i} = \frac{M_I}{\sum_{j\in\mathcal{I}_I}m_j} \sum_{i=1}^{n} \p_i, \qquad \textrm{for}~I =1,\ldots,N. 
\end{equation*}
When $M_I := \sum_{j \in \mathcal{I}_I} m_j$, it simplifies to $\mathcal{M}_{I}(\p^n) = \sum_{i=1}^{n} \p_i$. This condition, as will be demonstrated in Section~\ref{sec:sub:sub:momen}, corresponds to momentum space consistency.

With the mapping function defined, we can now analyze the relationship between the atomistic and CG scales within a statistical mechanics framework. Specifically, we consider the phase-space probability distributions of both models. A CG model is deemed consistent with its underlying atomistic model if the probability distribution of the CG model matches that of the CG sites derived from the atomistic model via the mapping function~\cite{noid2008multiscale}.

The phase-space probability distribution of the atomistic system is determined by its Hamiltonian $\mathcal{H}_{\rm AA}= \sum^n_{i=1} \p^2_i/(2m_i) + u(\r, \z)$. Then the phase-space probability distribution factorises into independent contributions from coordinates and momenta,
\begin{equation*}
    p_{rp}(\r, \p, \z) = p_{r}(\r, \z) \cdot p_p(\p) \propto e^{-u(\r, \z)/(k_BT)} \cdot e^{-\sum^n_{i=1} \p^2_i/(2m_i k_BT)},
\end{equation*}
where $p_{r}(\r, \z)$ and $p_{p}(\p)$ are the probability distributions of coordinates with atomic species and momenta, respectively (defined up to a normalisation factor). Analogously, the CG model is described by the Hamiltonian $\mathcal{H}_{\rm CG} = \sum^N_{I=1} \P^2_I/(2M_I) + U(\R, \bzeta)$,
and the probability distribution of the CG model factorises as
\begin{equation*}
    P_{RP}(\R, \P, \bzeta) = P_{R}(\R, \bzeta)\cdot P_P(\P) \propto e^{-U(\R, \bzeta)/(k_BT)} \cdot e^{-\sum^N_{I=1} \P^2_I/(2M_I k_B T)}.
\end{equation*}

We now make use of the mapping function and consider the probability of finding a given CG configuration within the atomistic ensemble, $p_R(\R)$. This is obtained by integrating out the internal degrees of freedom, averaging over their fluctuations for both positions and momenta:
\begin{eqnarray}\label{eq:pRpP}
    p_R(\R, \bzeta) &=& \int p_r(\r, \z) \prod_{I=1}^N \delta\big(\mathcal{M}_{I}(\r)-\R_I\big) {\rm d}\r, \\ 
    p_P(\P) &=& \int p_p(\p) \prod_{I=1}^N \delta\big(\mathcal{M}_{I}(\p)-\P_I\big) {\rm d}\p.
\end{eqnarray}

\subsubsection{Momentum space}
\label{sec:sub:sub:momen}

For consistency in momentum space we require
\begin{eqnarray}\label{eq:Pequiv}
    P_P(\P) = p_P(\P).
\end{eqnarray}
The left-hand side of this equation can be written as a product of independent zero-mean Gaussians for each $I$,
\begin{eqnarray}\label{eq:PP}
    P_P(\P) \propto \prod_{I=1}^N e^{-\P^2_I/(2M_I k_B T)}.
\end{eqnarray}
Taking into account \eqref{eq:pRpP}, \eqref{eq:Pequiv} with \eqref{eq:PP} imposes two conditions on the CG model: (i) for $p_P(\P)$ to be a product of independent zero-mean Gaussians as well, no atom can be involved in the definition of more than one CG site; and (ii) requiring the following condition on the CG masses $M_I$ for all CG site $I$
\begin{eqnarray*}
    \frac{1}{M_I} = \sum_{i\in\mathcal{I}_I} \frac{w^2_{Ii}}{m_i},
\end{eqnarray*}
with which the center of mass results in $M_I := \sum_{i\in\mathcal{I}_I}m_i$.

Note that consistency in momentum space only implies that the momentum distribution in the CG model matches the one in the atomistic model. Next, we focus on the consistency in configuration space.

\subsubsection{Configuration space}

Analogously, we consider the CG model to be consistent with the underlying atomistic model in configuration space. To that end, one requires
\begin{eqnarray*}
    P_R(\R) = p_R(\R),
\end{eqnarray*}
which is equivalent to
\begin{eqnarray*}
    e^{-U(\R, \bzeta)/(k_B T)} \propto \int e^{-u(\r, \z)/(k_B T)}\prod_{I=1}^N \delta\big(\mathcal{M}_{\R_I}(\r)-\R_I\big) {\rm d}\r,
\end{eqnarray*}
thereby defining the {\it consistent} CG potential $U(\R, \bzeta)$. Note that this is not a potential energy, but also contains an entropic contribution due to integrating over fluctuations in the internal degrees of freedom. Hence, this CG potential actually describes a free energy surface, and it explicitly depends on the temperature $T$. We can explicitly write $U$ as a {\it conditional free energy}, or the so-called potential of mean force (PMF):
\begin{eqnarray}\label{eq:many-body-PMF}
    U(\R, \bzeta) = -k_B T \log Z(\R, \bzeta) + c,
\end{eqnarray}
where
\begin{eqnarray}\label{eq:ZR}
    Z(\R, \bzeta) = \int z(\r, \z) \prod^N_{I=1}\delta\big(\mathcal{M}_I(\r)-\R_I\big) {\rm d}\r
\end{eqnarray}
is the constrained partition function (as a function of the CG coordinates). When the CG interactions are given by the PMF for the CG coordinates, both structural and thermodynamic properties of the atomistic model are exactly preserved. 

\subsection{Mean force and instantaneous collective force}

We start from the {\it consistent} CG potential based on the many-body PMF. The coarse-grained force on site $I$ is given by the negative gradient of the PMF, $\F_I(\R, \bzeta) = -\nabla_{\R_I} U(\R, \bzeta)$. Carrying out the derivative, we obtain
\begin{eqnarray}\label{eq:F}
    \F_I(\, \bzeta) = \frac{k_B T}{Z(\R, \bzeta)} \int e^{-u(\r, \z)/(k_B T)} \prod_{J (\neq I)} \delta \big(\mathcal{M}_{J}(\r)-\R_J\big)\frac{\partial}{\partial \R_I} \delta \big(\sum_{i\in\mathcal{I}_I}w_{Ii}\r_i-\R_I\big) {\rm d}\r.
\end{eqnarray}
We integrate this by parts, making use of the equation
\begin{eqnarray*}
    \frac{\partial}{\partial \R_I} \delta \big(\sum_{i\in\mathcal{I}_I}w_{Ii}\r_i-\R_I\big) = -\frac{1}{w_{Ik}} \frac{\partial}{\partial \r_k} \delta \big(\sum_{i\in\mathcal{I}_I}w_{Ii}\r_i-\R_I\big),
\end{eqnarray*}
where we chose an arbitrary atom $k$ involved in the definition of CG site $I$. Let the force on atom $j$ in the atomistic system be $\f_j(\r, \z) = - \nabla_{\r_j} u(\r, \z)$ and the instantaneous collective force (ICF) is defined by $\mathcal{F}_I(\r, \z) = \sum_{j\in\mathcal{I}_I} \f_j(\r, \z)$. Integration of \eqref{eq:F} is then straightforward and yields
\begin{eqnarray}\label{eq:MF}
    \F_{I}(\R, \bzeta) = \Bigg\<\frac{\int e^{-u(\r, \z)/(k_B T)}\prod_{J}\delta\big(\mathcal{M}_J(\r)-\R_J\big) \mathcal{F}_I(\r, \z) {\rm d}\r  }{\int e^{-u(\r, \z)/(k_B T)}\prod_J\delta\big(\mathcal{M}_J(\r)-\R_J\big) {\rm d}\r } \Bigg\>_{\R},
\end{eqnarray}
which is the conditional expectation value (or equilibrium average) in the ensemble of the atomistic system; this expectation value is a function of the CG coordinates $\R$ (which are kept constant while averaging). $\bzeta = \bzeta(\zz)$ is determined by the microscopic configuration for which we assume in this integral that $\zz$ remains constant. In practice, this ensemble average will be replaced by a time average over atomistic trajectories.

The ICF can be considered noisy samples of the mean forces. The optimal CG forces, correspond to a {\it consistent} CG model, are derived from the atomistic forces through an equilibrium average over the internal degrees of freedom, with the CG coordinates kept constant. This describes the mean forces on the CG sites. If the CG force field has the flexibility to reproduce this, which will be the case for an $N$-body potential, and the CG model will be {\it consistent} with the reference all-atom model.

\subsection{Iterative Boltzmann inversion}
\label{sec:apd:sub:IBI}

Iterative Boltzmann Inversion (IBI)~\cite{moore2014derivation} is a structure-based coarse-graining method that derives coarse-grained interactions from the structural distributions of the underlying atomistic system. The approach relies on one-dimensional distributions of collective variables (CVs), such as the pairwise distance $r$, to determine the effective interactions.

Iterative Boltzmann Inversion (IBI) refines the pair potentials obtained from Direct Boltzmann Inversion (DBI) to reproduce atomistic distributions more accurately. While applicable to various degrees of freedom, we focus on non-bonded pair potentials and the RDFs. Starting with an initial guess $U^{(0)}(r)$, often derived from DBI, IBI iteratively updates the CG pair potential as follows:
\[
U^{(i+1)}(r) = U^{(i)}(r) + \alpha \Delta U^{(i)}(r),
\]
where $\alpha$ ($0 < \alpha \leq 1$) is an optional scaling factor to stabilize convergence. In this work, we set $\alpha = 0.2$. The correction term $\Delta U^{(i)}(\r)$ is defined as:
\[
\Delta U^{(i)}(r) = k_B T \log\left(\frac{g^{(i)}_{\rm CG}(r)}{g_{\rm AA}(r)}\right),
\]
where $g_{\rm CG}^{(i)}(r)$ and $g_{\rm AA}(r)$ are the RDFs obtained from the CG potential $U^{(i)}(r)$ and atomistic models, respectively. The iterations continue until the CG RDF sufficiently matches the atomistic RDF.

This scheme is motivated by the following observation: when $g^{(i)}_{{\rm CG}}(r) > g_{\rm AA}(r)$ at
a given pair distance $r$, this implies that the PMF of the CG model at that distance is
too attractive, and that it should be more repulsive. As the PMF is given by the sum
of direct pair interaction $U^{(i)}(r)$ and environmental contributions, the simplest way to
increase the PMF is to increase the value of the pair potential at that distance. Clearly, $g_{\rm CG}^{*}(r)=g_{\rm AA}(r)$ is a fixed point of this iteration. Hence, assuming convergence, the IBI procedure results in a pair potential that reproduces the RDF of the all-atom system.

\subsection{Atomic Cluster Expansion}
\label{sec:apd:ACE}

We provide a brief overview of the Atomic Cluster Expansion (ACE) potential and refer to~\cite{2019-ship1, 2021-qmmm3, wang2020posteriori, witt2023otentials} for more detailed construction and discussion. The first step is to decompose the total potential into site potentials, 
\begin{eqnarray}\label{eq:acecg_local_site_appendix}
    U^{\rm ACE}(\bth; \R, \bzeta) = \sum^{N}_{I=1} U^{\rm ACE}_{I}(\bth; \R, \bzeta).
\end{eqnarray}
The definition of $U^{\rm ACE}_{I}$ involves specifying a cutoff radius $R_{\rm cut}$ so that only CG particles $\R_J$ within distance $R_{IJ} < R_{\rm cut}$ contribute to the site energy $U^{\rm ACE}_{I}$.

The second step is to express each site-potential in terms of a truncated {\it many-body expansion}~\cite{2019-ship1, Drautz19},  
\begin{eqnarray}\label{eq:acecg_local_appendix}
    U^{\rm ACE}_I(\bth; \R, \bzeta) = \sum_{\nu = 0}^{\nu_{\rm max}} 
        \sum_{1 \leq I_1 < \dots < I_\nu \leq N}
        W^{(\nu)}\big(\bth; \zeta_I, \R_{I_1}, \zeta_{I_1}, \dots, \R_{I_\nu}, \zeta_{I_\nu}). 
\end{eqnarray}
Each potential $W^{(\nu)}$ is then expanded in terms of a tensor product basis~\cite{2019-ship1},
\begin{equation*}
\phi_{{\z} {\bm n} {\bm \ell} {\bm m}}(\R, \bzeta) := \prod_{\alpha=1}^{\nu} \phi_{{\z_{\alpha} {\bm n}_{\alpha} {\bm \ell}_{\alpha} {\bm m}_{\alpha}}}(\R_{II_{\alpha}}, \zeta_{I_{\alpha}}),
\end{equation*}
where the one-particle basis $\phi_{z n \ell m}(\mathbf{r}, \zeta) := \delta_{z\zeta} P_n(r)Y_{{\bm \ell}}^{{\bm m}}(\hat{\mathbf{r}}),$ with $\mathbf{r} \in \mathbb{R}^d$, $r = |\mathbf{r}|$, and $\hat{\mathbf{r}} = \mathbf{r}/r$. The functions $Y_{\ell}^{m}$ denote the complex spherical harmonics for $\ell = 0,1,\ldots$, and $m=-\ell,\ldots,\ell$, and $P_n(r)$ are radial basis functions for $n=0,1,\ldots$. 

This parameterization is already invariant under permutations (or, relabelling) of the input structure $(\R, \bzeta)$. One then symmetrizes the tensor product basis with respect to the group $O(3)$, employing the representation of that group in the spherical harmonics basis. This results in a linear parameterization
\begin{align*}
    U^{\text{ACE}, (\nu)}_{I}(\R, \bzeta)
    &= \sum_{{\z{\bm n} {\bm \ell} q}} c_{{\z{\bm n} {\bm \ell} q}} B_{\z{\bm n} {\bm \ell} q} (\R, \bzeta), \\ 
    \text{where} 
    \quad     
    B_{\z{\bm n} {\bm \ell} q} (\R, \bzeta)
    &= \sum_{{\bm m } \in \mathcal{M}_{ {\bm \ell}}} \mathcal{C}^{{\bm n} {\bm \ell} q}_{{\bm m }} {\bm A}_{\z{\bm n} {\bm \ell}{\bm m} }(\R, \bzeta), \\ 
    \text{and} 
    \quad 
    {\bm A}_{\z{\bm n} {\bm \ell}{\bm m} }(\R, \bzeta) &= \prod_{\alpha=1}^{\nu} \sum_{j=1}^{J} \phi_{\z_{\alpha}{\bm n}_{\alpha}{\bm \ell}_{\alpha}{\bm m}_{\alpha}}(\R_{II_{j}}, \zeta_{I_{j}}).
\end{align*}
The $q$-index ranges from $1$ to $n_{\bm n \bm \ell}$ and enumerates the number of all possible invariant couplings through the generalized Clebsch--Gordan coefficients $\mathcal{C}^{{\bm n} {\bm \ell} q}_{{\bm m }}$. This parameterization was originally devised in \cite{Drautz19}. Its approximation properties and computational complexity are analyzed in detail in \cite{2019-ship1, 2021-apxsym, PhysRevMaterials.6.013804, braun2022higher}. The implementation we employ in our work is described in \cite{witt2023otentials}. The latter reference also described the details of the choice of radial basis $P_n$.

To complete the description of our parameterization, we select a finite subset of the basis, 
\begin{align*}
\mathbf{B} := \Big\{ B_{\z{\bm n} {\bm \ell} q} \,\Big|\, & ({\bm n}, {\bm \ell}) \in \mathbb{N}^{2N} \text{ ordered}, ~ \sum_{\alpha} \ell_{\alpha} \text{ even},  
    \sum_{\alpha} m_\alpha = 0, \\ 
    & q = 1, \ldots, n_{{\bm n}{\bm \ell}}, 
    \sum_\alpha \ell_\alpha + n_\alpha \leq D_{\rm tot}, 
    N \leq \mathscr{N} 
    \Big\},
\end{align*}
where the two approximation parameters are the correlation order $\nu_{\rm max}$ and the total degree $D_{\rm tot}$. With this selection of the basis we can write our ACE parameterization more convenient as 
%
\begin{equation}
U^{\text{ACE}}(\bth; \R, \bzeta) = \sum_{B \in \mathbf{B}} \theta_B B(\R, \bzeta), 
\end{equation}
where $\bth = (\theta_B)_{B \in {\bf B}}$ are the model parameters. Each parameter $\theta_B$ can be directly interpreted as a polynomial coefficient for one of the potentials $W^{(\nu)}$.
%
Due to the completeness of this representation, as the approximation parameters (like body-order, cut-off radius, and expansion precision) approach infinity, the model can represent any arbitrary potential \cite{2019-ship1}.

\section{Simulation supplement}
\label{sec:apd:ACE:sub:hyperparameters}

In this section, we provide supplementary details on the simulation setup, focusing primarily on the hyperparameters used for training the ACE-CG models and performing MD simulations. 

The training of the ACE-CG models followed the architecture and methodology outlined in~\cite{witt2023otentials}. Model fitting was conducted using generalized Tikhonov regularization, where the Tikhonov matrix was constructed with an algebraic smoothness prior of strength $p=4$, as defined in~\cite[Eq.10]{witt2023otentials}. To estimate the scaling parameter in the generalized Tikhonov regularization, the Bayesian Linear Regression (BLR) solver was utilized, ensuring robust and reliable parameter fitting. The hyperparameters for training the models on polymer and methanol fluids are summarized in Table~\ref{tab:system_parameters}. In the table, $\nu_{\rm max}$ represents the correlation order (with body-order defined as $\nu_{\rm max}+1$, $D_{\rm max}$ indicates the polynomial degree, and $r_{\rm cut}$ denotes the cutoff radius. 

The MD simulations were performed in the canonical (NVT) ensemble using Langevin dynamics to maintain a constant temperature. The friction coefficient $\gamma$ was carefully chosen to ensure effective thermostating without overdamping low-frequency modes, with $\gamma$ proportional to the mass of each particle. For methanol fluids, $\gamma = 2 \, \text{ps}^{-1}$ was used, while for star-polymer fluids, $\gamma = 5 \, \text{ps}^{-1}$.

\begin{table}[h!]
\centering
\begin{tabular}{@{}ccccc@{}}
\toprule
\textbf{Systems}       & \textbf{Models} & $\nu_{\rm max}$ & $D_{\rm max}$ & $r_{\rm cut}$ \\ \midrule
\multirow{3}{*}{Polymer} & ACE-CG (BO2)        & 1             & 22         & 16.0                 \\ \cmidrule(l){2-5} 
                              & ACE-CG (BO3)        & 2             & 18         & 16.0                            \\ \cmidrule(l){2-5} 
                              & ACE-CG (BO4)        & 3             & 16         & 16.0                            \\ \midrule
\multirow{3}{*}{Methanol (one-site)}     & ACE-CG (BO2)        & 1             & 18         & 7.0                            \\ \cmidrule(l){2-5} 
                              & ACE-CG (BO3)        & 2             & 14         & 7.0                            \\ \cmidrule(l){2-5} 
                              & ACE-CG (BO4)        & 3             & 14         & 7.0                           \\ \midrule
\multirow{3}{*}{Methanol (two-sites)}     & ACE-CG (BO2)        & 1             & 14         & 6.0                            \\ \cmidrule(l){2-5} 
                              & ACE-CG (BO3)        & 2             & 12         & 6.0                           \\ \cmidrule(l){2-5} 
                              & ACE-CG (BO4)        & 3             & 12         & 6.0                          \\ \bottomrule
\end{tabular}
\caption{Summary of hyperparameters for training ACE-CG models for polymer and methanol fluids.}
\label{tab:system_parameters}
\end{table} 

\section{Further results}
\label{sec:apd:numerics}

In this section, we present additional numerical results to support and maintain the findings and conclusions discussed in the main text. 

\begin{figure}[htb]
\begin{center}
\includegraphics[height=6.0cm]{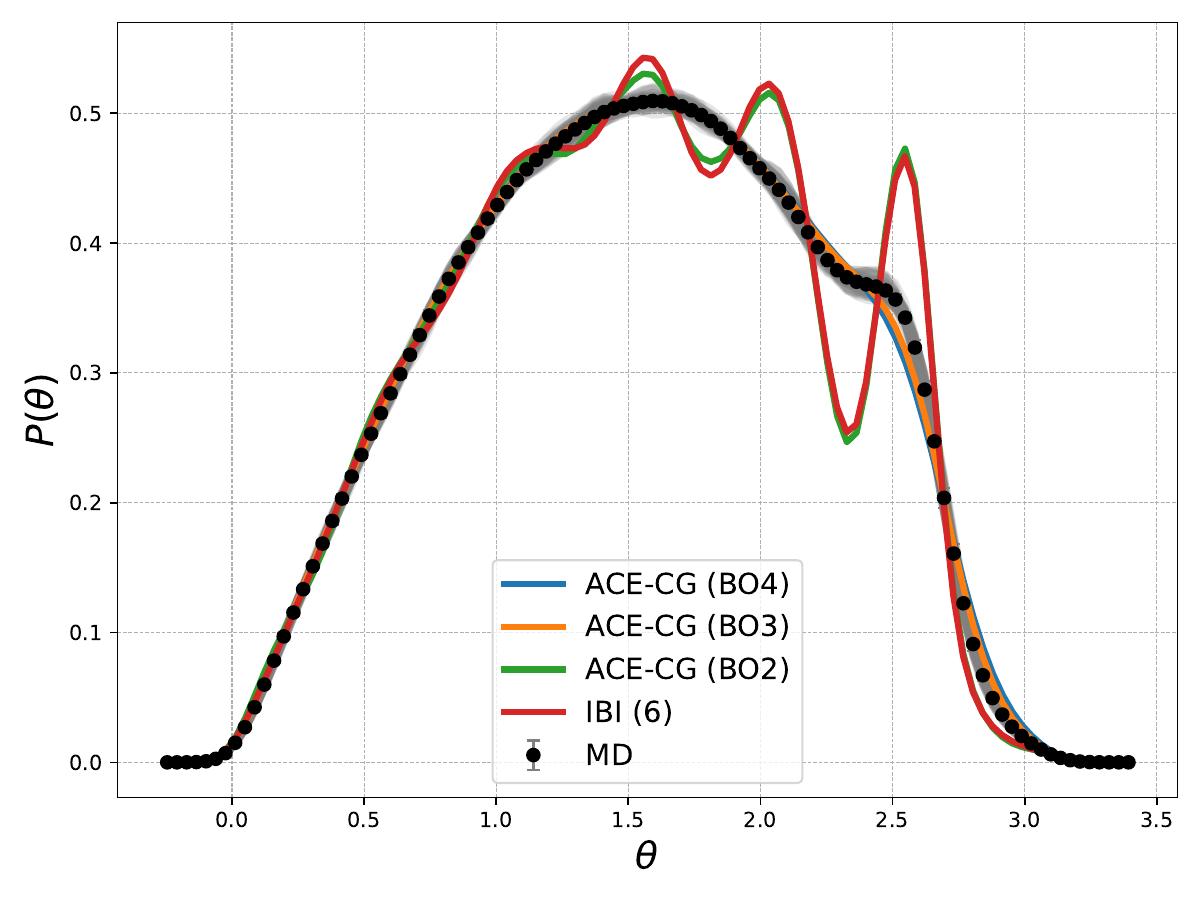}
	\caption{Polymer fluids: ADF with $r_{\rm cut}=30\mathring{\textrm{A}}$.} 
	\label{figs:star_poly_ADF_appendix}
\end{center}
\end{figure}

Figure~\ref{figs:star_poly_ADF_appendix} illustrates the ADF for polymer fluids with $r_{\rm cut}=30\mathring{\textrm{A}}$. It is evident that both the IBI and pairwise interaction CG models perform poorly, showing significant deviations from the MD reference results. In contrast, ACE-CG models with body orders 3 and 4 align closely with the MD reference, highlighting once again the critical role of many-body interactions in accurately capturing the behavior of polymer fluid systems.

\begin{figure}[htb]
\begin{center}
\includegraphics[height=6.0cm]{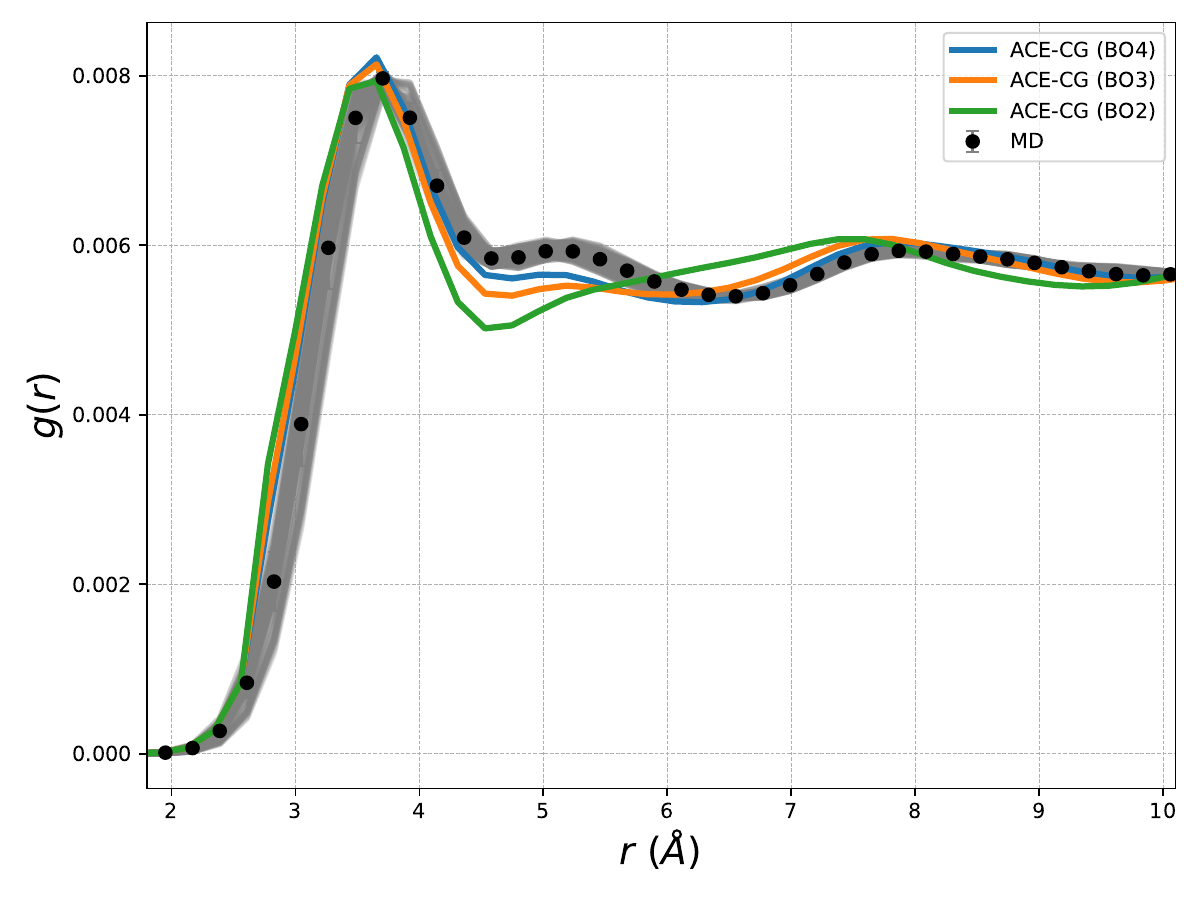}
\caption{Methanol fluids: RDF for mixed carbon-oxygen group.}
	\label{figs:pure_methanol_RDF_2_appendix}
\end{center}
\end{figure}

Figure~\ref{figs:pure_methanol_RDF_2_appendix} presents the mixed carbon-oxygen RDFs. The results demonstrate excellent agreement with the reference MD data, particularly for higher body order cases, underscoring the accuracy of the ACE-CG models in capturing interspecies interactions.

\begin{figure}[htb]
    \centering
    \begin{subfigure}[b]{0.32\textwidth}
        \centering
        \includegraphics[height=4.1cm]{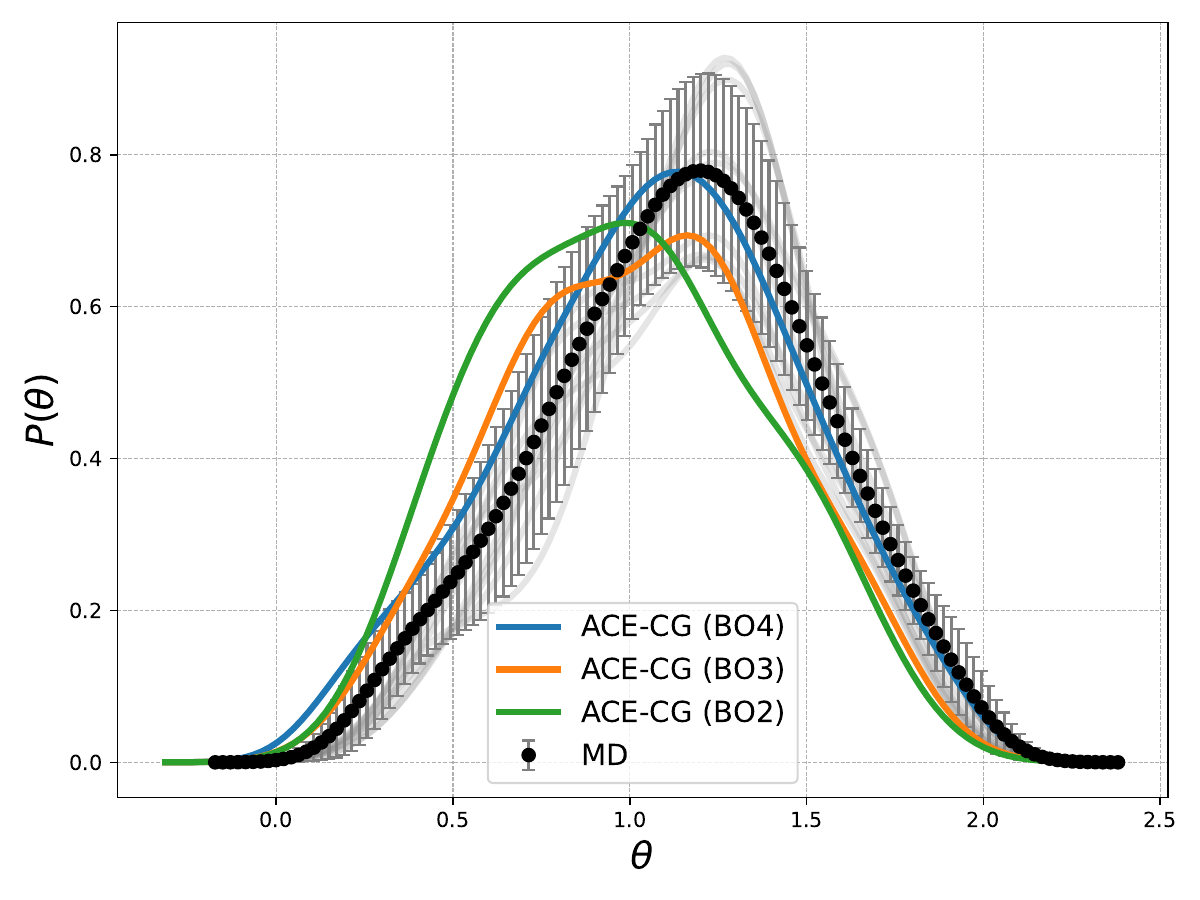}
        \caption{$r_{\rm cut} = 4.5 \mathring{\textrm{A}}$}
        \label{fig:adf_local_2_4.5_O}
    \end{subfigure}
    \begin{subfigure}[b]{0.32\textwidth}
        \centering
        \includegraphics[height=4.1cm]{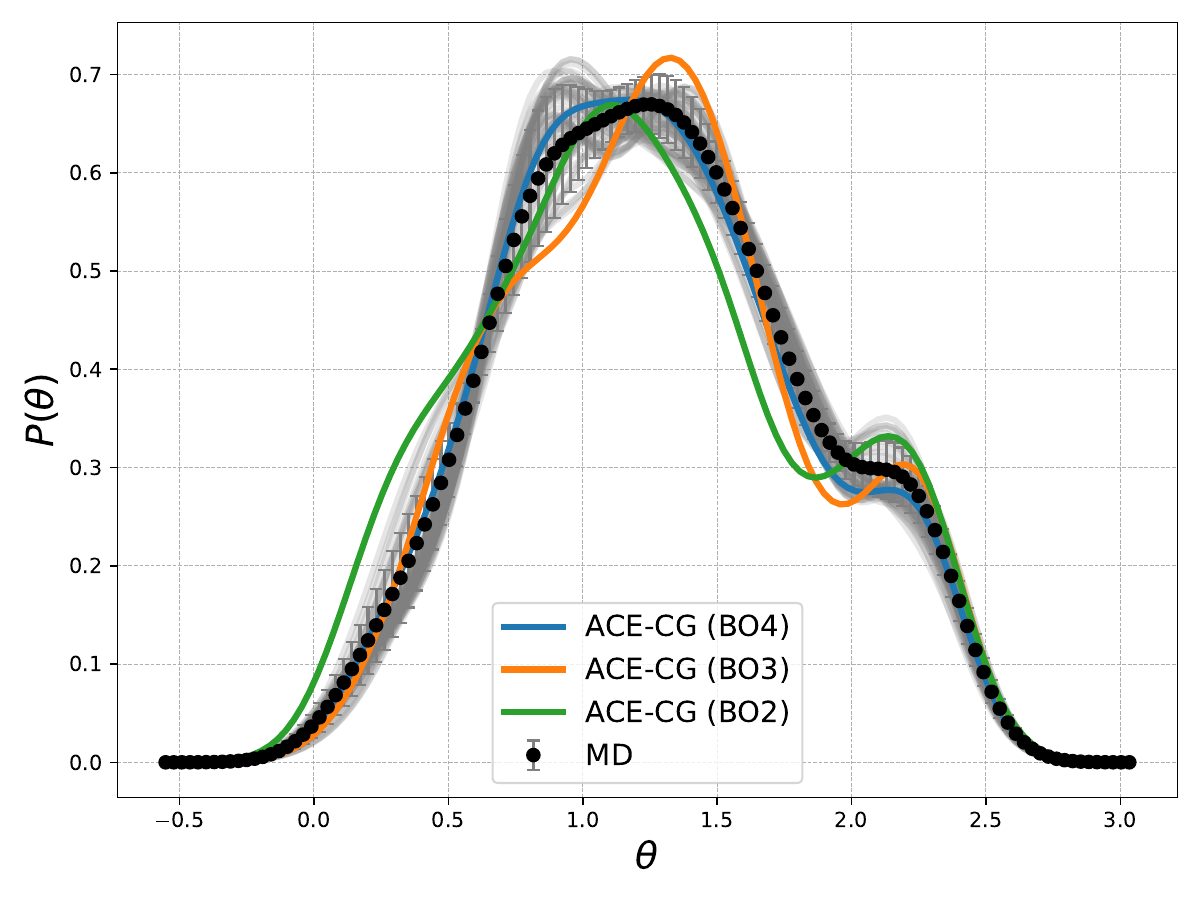}
        \caption{$r_{\rm cut} = 6.0 \mathring{\textrm{A}}$}
        \label{fig:adf_local_2_6.0_O}
    \end{subfigure}
    \begin{subfigure}[b]{0.32\textwidth}
        \centering
        \includegraphics[height=4.1cm]{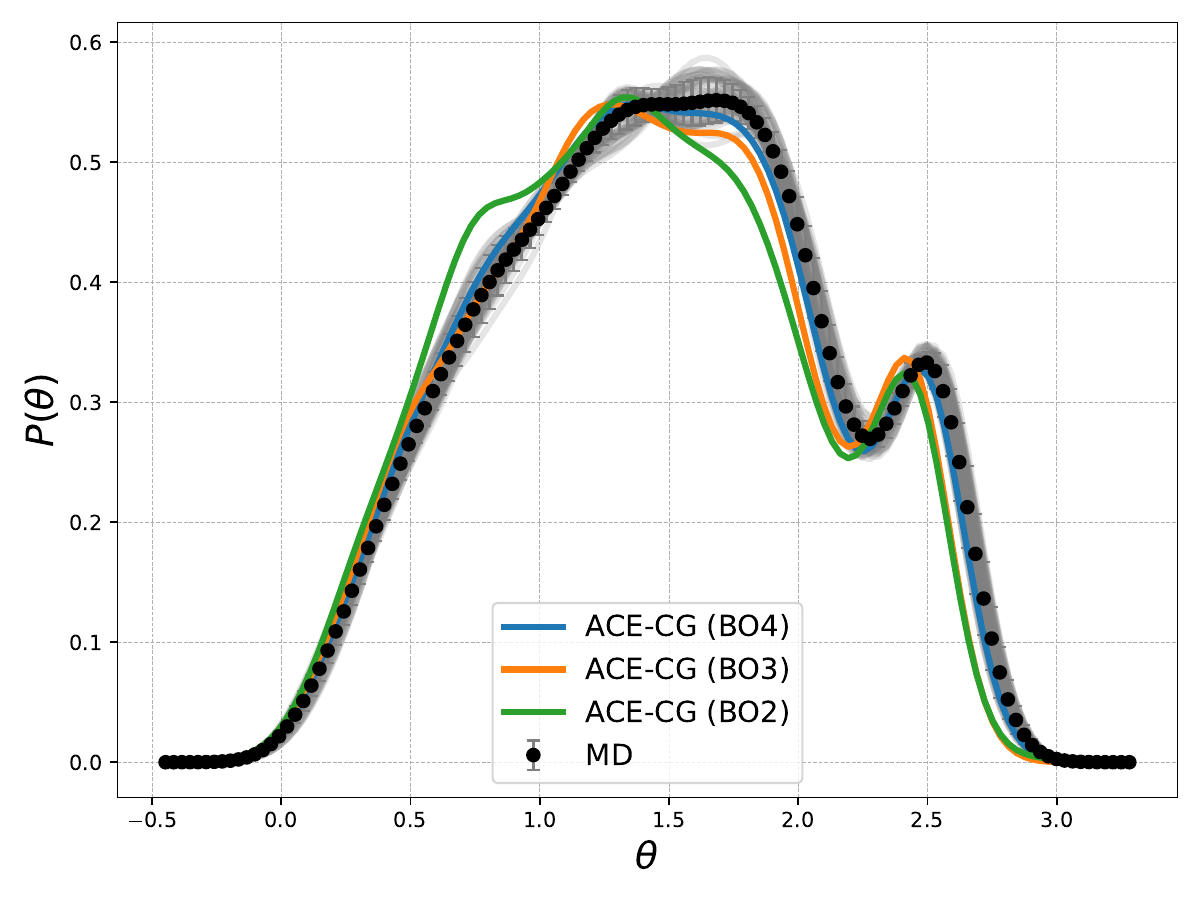}
        \caption{$r_{\rm cut} = 8.0 \mathring{\textrm{A}}$}
        \label{fig:adf_local_2_8.0_O}
    \end{subfigure}
    \caption{Methanol fluids: two-sites ADF for oxygen group with different cutoff radii.}
    \label{figs:pure_methanol_ADF_2_OOO_appendix}
\end{figure}

Figure~\ref{figs:pure_methanol_ADF_2_OOO_appendix} displays the two-site ADFs for the oxygen group under varying cutoff radii. While the smallest cutoff radius exhibits a relatively larger deviation, the ACE-CG models still achieve qualitatively accurate results, maintaining the overall shape consistent with the MD reference. Notably, ACE-CG with four-body interactions delivers the most accurate predictions, aligning well with our previous observations.

\bibliographystyle{plain}
\bibliography{cg.bib}

\end{document}